\documentclass[12pt,preprint]{aastex}

\usepackage{natbib}

\slugcomment{40 pages, 4 tables, 8 figures; For submission to \apj}

\shorttitle{\object{Kapteyn's Star}: Rotation, X-ray and UV}
\shortauthors{Guinan et al.}

\begin{document}


\title{Living with a Red Dwarf: \\
Rotation and X-ray and Ultraviolet Properties of the Halo Population \object{Kapteyn's Star} \footnote{Based on observations made with the NASA/ESA \textit{Hubble Space Telescope}, obtained at the Space Telescope Science Institute, which is operated by the Association of Universities for Research in Astronomy, Inc., under NASA contract NAS 5-26555. These observations are associated with program \#13020. This work is also based on observations obtained with the \textit{Chandra X-ray Observatory}, a NASA science mission, program \#13200633.}}


\author{Edward F. Guinan, Scott G. Engle and Allyn Durbin}
\affil{Department of Astrophysics and Planetary Science, Villanova University,
    Villanova, PA 19085 USA}
\email{scott.engle@villanova.edu}



\begin{abstract}
As part of Villanova's \textit{Living with a Red Dwarf} program, we have obtained UV, X-ray and optical data of the Population II red dwarf -- \object{Kapteyn's Star}. \object{Kapteyn's Star} is noteworthy for its large proper motions and high RV of $\sim+245$ km s$^{-1}$. As the nearest Pop II red dwarf, it serves as an old age anchor for calibrating Activity/Irradiance-Rotation-Age relations, and an important test bed for stellar dynamos and the resulting X-ray -- UV emissions of slowly rotating, near-fully convective red dwarf stars. Adding to the notoriety, \object{Kapteyn's Star} has recently been reported to host two super-Earth candidates, one of which (Kapteyn b) is orbiting within the habitable zone \citep{ang14,ang15}. However, Robertson et al. (2015) questioned the planet's existence since its orbital period may be an artifact of activity, related to the star's rotation period. Because of its large Doppler-shift, measures of the important, chromospheric H \textsc{i} Lyman-$\alpha$ 1215.67\AA~emission line can be reliably made, because it is mostly displaced from ISM and geo-coronal sources. Lyman-$\alpha$ emission dominates the FUV region of cool stars. Our measures can help determine the X-ray--UV effects on planets hosted by \object{Kapteyn's Star}, and planets hosted by other old red dwarfs. Stellar X-ray and Lyman-$\alpha$ emissions have strong influences on the heating and ionization of upper planetary atmospheres and can (with stellar winds and flares) erode or even eliminate planetary atmospheres. Using our program stars, we have reconstructed the past exposures of \object{Kapteyn's Star}'s planets to coronal -- chromospheric XUV emissions over time.
\end{abstract}





\section{Introduction}

Red dwarfs (dwarf-M stars = dM stars) comprise about 75\% of the local stellar population \citep{hen16}. Recent statistical analyses of \textit{Kepler} Mission exoplanet data indicate that a significant fraction ($\sim$15 -- 25\%) of red dwarfs are expected to host terrestrial-size planets within their habitable zones \citep[see: e.g.][]{dre13,dre15,bon13}. This indicates statistically that $\sim$35 -- 50 potentially habitable, terrestrial-size planets are expected to orbit the 240+ red dwarfs within $\sim$10 pc ($\sim$33 light years) of the Sun. 

Typically, these low mass, low luminosity, cool stars have very slow nuclear fusion rates and thus have very long ($>100$ Gyr) main-sequence lifetimes. Because of their slow nuclear evolution, the measurable age-dependent properties of red dwarfs, such as luminosity, radius, and temperature, change very slowly so that the isochronal stellar evolution models employed for higher mass stars are not applicable. This means that asteroseismic age determinations (like those being used to determine ages and masses for (mostly) F--G stars from the \textit{Kepler} Mission -- see \citealt[and references therein]{cha15,sil15}) of red dwarfs are also not feasible. Instead, other less direct methods have been developed for estimating the ages of red dwarfs. Like the Sun and solar-type stars, red dwarfs lose angular momentum with time, resulting in the lengthening of rotation periods with age \citep[e.g.][]{eng11}. This occurs from magnetic braking in which wind-embedded magnetic fields carry angular momentum away from the star. Villanova's \textit{Living with a Red Dwarf} program is focusing on using the stellar rotation rates, as well as rotation-related magnetic dynamo activity tracers, such as coronal X-ray, transition region / chromospheric FUV-UV and H \textsc{i} Lyman-$\alpha$ 1215.67\AA~ emission (hereafter Ly$\alpha$), to provide relations for reliable age determinations. Because of the high frequency and long lifetimes of red dwarfs, there could be a greater possibility of more advanced life (even intelligent extraterrestrial life) in solar-age (or older) red dwarf exoplanet systems. \textit{MEarth} \citep[][and references therein]{irw15}, \textit{UVES} \citep[][and references therein]{tuo14}, \textit{HARPS} \citep[][and references therein]{ast15}, \textit{SDSS-III} \citep[][and references therein]{des13}, and the upcoming \textit{TESS} \citep{pep15} mission are some surveys that are targeting red dwarfs in the search for orbiting, potentially habitable planets. Also, the proximity of a large number of red dwarfs with planets (including Kapteyn's Star) makes them compelling targets for the upcoming \textit{James Webb Space Telescope} (\textit{JWST}) to secure spectra and search for bio-signatures for possible life.

As part of the \textit{Living with a Red Dwarf} program, we have obtained \textit{Hubble Space Telescope Cosmic Origins Spectrograph} (\textit{HST-COS}) UV spectra and \textit{Chandra} \textit{Advanced CCD Imaging Spectrometer} (\textit{ACIS-I}) X-ray observations of numerous red dwarfs, including the bright ($V \approx 8.9$), nearby (3.91 pc), early M dwarf/subdwarf -- \object{Kapteyn's Star} (see Table 1 for select stellar properties). Over the past decade we have also been carrying out extensive CCD photometry to investigate star spot properties and measure rotation periods. These measurements are crucial to the study of red dwarfs, but particularly those with advanced ages, because \object{Kapteyn's Star} is the nearest halo star (with an estimated age of $\sim11.5^{+0.5}_{-1.5}$ Gyr -- see Table 1). Recent studies of rotation-age-activity relations for red dwarfs have been carried out by \citet{kir07,rei08,eng11} and by \citet{wes15}. Although these studies utilize different data sets and methods, they generally agree that red dwarfs (earlier than $\sim$M5--6) decrease in magnetic activity with increasing age and increasing (longer) rotation periods, as defined by H$\alpha$ emission, starspot coverage, stellar flare rates or coronal X-ray emission. In our program, \object{Kapteyn's Star} is one of the oldest (if not \textit{the} oldest) red dwarf and, as such, serves as an anchor for our activity-rotation-age relations. Our observations permit a calibration at the old age / slow rotation / low activity extremes for the relations, and provide insights into the coronal X-ray and UV (XUV) emissions of old, slowly rotating dM stars. These results have been reported (in part) previously by \citet{dur14} and \citet{gui15}. 

\object{Kapteyn's Star} also provides crucial data for determining the XUV irradiances of habitable zone (HZ) planets hosted by older red dwarfs. Planetary atmospheres can be profoundly affected by the XUV radiation from their host stars -- especially when the stars are young and very active. This radiation is important for understanding the long-term habitability of planets \citep{rib05,buc06,fra15}. It also helps drive atmospheric dynamics by influencing ionosphere temperature profiles and density, in some cases causing potentially significant, rapid atmospheric expansion and mass escape \citep{gri04,lam08}. UV radiation produces photochemical reactions, some of which can lead to possible pre-biological molecules, molecular oxygen and ozone \citep{tia13,dom12}. The determination of stellar XUV radiation is critically important because it contains the strong chromospheric H \textsc{i} Ly$\alpha$ emission line -- by far the strongest Far-UV (FUV) emission line. However, only a small number of dM stars have been studied in the FUV \citep[see][and references therein]{lin14}, despite the critical importance that this radiation plays in the evolution of hosted planetary atmospheres and their potential for life. More recently, several additional, nearby red dwarfs have been observed with the \textit{Hubble Space Telescope} as part of the\textit{MUSCLES HST Heritage Survey} \citep[see][]{fra13,you16}. The XUV observations of \object{Kapteyn's Star} presented here provide the \textit{Living with a Red Dwarf} program with important data for the oldest dM stars. 

As an added bonus, \citet{ang14} have recently reported two super Earth-size planets orbiting \object{Kapteyn's Star} (our study of \object{Kapteyn's Star} began a few years prior to this discovery). One of these planets (Kapteyn b: $P_{\rm orb}$ = 48.6 days, $a$ = 0.168 AU; $M_{\rm P} \geq 4.8\pm1 M_{\oplus}$) orbits within the star's habitable zone. If confirmed, it would be the oldest terrestrial-type planet found so far, and inform the number and likelihood of planets that can form in the metal-poor environments of Pop II stars. However, \citet{rob15} question the existence of Kapteyn b. They claim that the small radial velocity (RV) signal (attributed to the gravitational effects of this planet on the host star) instead arises from the rotation-activity effects of the star. Subsequently, as discussed in more detail later, \citet{ang15} defend their initial claim.

Because of the high frequency of older red dwarfs, and their longevities ($> 50$ Gyr), red dwarf planetary systems may have a greater possibility to develop more advanced life than what has developed on Earth. The study of these stars, along with the determination of calibrated age relations, provides the dynamic evolutionary history of planetary systems, and an assessment of their stability with specific attention towards exoplanetary atmospheres and habitability. 

This paper is organized as follows: Section 2 is a brief discussion of \object{Kapteyn's Star}, including origin and age and its physical properties, along with a brief discussion of its possible hosted planets. The determination of the rotation of \object{Kapteyn's Star} from photometry and archival spectra are presented in Section 3. Section 4 includes the presentation and analysis of X-ray data from \textit{Chandra}, including \object{Kapteyn's Star}'s location in our red dwarf X-ray Activity-Age relation. In Sections 5, UV spectra obtained with \textit{COS} are presented, along with a Ly$\alpha$ Activity-Age relation using the newly obtained data for \object{Kapteyn's Star}. Section 6 provides a discussion and summary of our results including their impact on planetary habitability. Conclusions, as well as future plans and expectations are provided in Section 7.

\section{\protect\object{Kapteyn's Star} and its Planetary System}

\subsection{Characteristics of an Old Red Dwarf}

\object{Kapteyn's Star} (GJ 191; HD 33793) attained notoriety over a century ago when the star was discovered by \citet{kap97} to have very high proper motions ($\mu_{\alpha}$ = +6505.08 mas yr$^{-1}$;  $\mu_{\delta}$ = 5730.84 mas $^{-1}$ \citep{van07}. \citet{cur09} measured its space motions and, until 1916 (when displaced by Barnard's Star), it had the highest known space velocity relative to the Sun \citep[see][]{kot05}. This high relative velocity indicated that it was a Halo (Pop II) star, and subsequent studies have shown that \object{Kapteyn's Star} is indeed metal-poor, with $[Fe/H] = -0.86\pm0.05$ \citep[][and references therein]{woolf05}. Modern measures of its proper motions from \textit{Hipparcos} \citep{van07} and radial velocity of $RV = +245.2$ km s$^{-1}$ \citep[from][]{nid02} yield very high (\textit{U}, \textit{V}, \textit{W}) space motions of $+21.1\pm0.3, -287.8\pm0.3, -52.6\pm0.3$ km s$^{-1}$ \citep{kot05}. \citet{egg96} identified a number of other stars with similar space motions to \object{Kapteyn's Star}. He proposed these stars as members of the Kapteyn Moving Group (Star Stream). These motions place \object{Kapteyn's Star} in an eccentric, retrograde galactic orbit. Subsequent studies of its space motions and low metal abundance indicate that \object{Kapteyn's Star} (and other possible moving group stars) may have originated from the old globular cluster $\omega$ Cen \citep{wyl10}. But the association of the Kapteyn Moving Group (though not \object{Kapteyn's Star} itself) with $\omega$ Cen has been recently questioned by \citet{nav15} on chemical abundance grounds. It is believed that $\omega$ Cen may be the remnant of an old dwarf spheroidal galaxy that interacted with our Galaxy \citep{bek03}. For the purposes of this study, we adopt an age of $11.5^{+0.5}_{-1.5}$ Gyr which is near the estimated mean age determinations of the $\omega$ Cen cluster \citep{kot05}. This age is also consistent with \object{Kapteyn's Star}'s combination of high space motions and low metallicity , placing it within the Halo population of the Milky Way, which has a recent age determination of $11.4\pm0.7$ Gyr \citep{kal12}. We note that low metallicity on its own would not be hard evidence of Halo membership but, combined with the space motions, it offers further support. The star's primary properties are summarized in Table 1.

\subsection{The \protect\object{Kapteyn's Star} Planetary System -- ``To be or not Kapteyn b''}

Until recently, the ``claim to fame'' of \object{Kapteyn's Star} was being one of the brightest and fastest moving red dwarfs, and the closest bona fide Pop II dM stars. Recently, though, \citet{ang14} reported that \object{Kapteyn's Star} may host two super-Earth planets. These planets were discovered via spectroscopic radial velocities. The measured $RV$ amplitudes are $K_{\rm b, c} = 2.25\pm0.31$ m s$^{-1}$ and $2.27\pm0.28$ m s$^{-1}$, respectively, for Kapteyn b and c. Kapteyn b has a minimum mass of $M_p~sin~i = 4.8\pm0.9 M_{\oplus}$, an orbital period of $P_{\rm orb} = 48.6$ days and a semi-major axis of $a = 0.168$ AU. Kapteyn c has a minimum mass of $M_p~sin~i \geq 7.0\pm1.1 M_{\oplus}$, orbits with a period of $P_{\rm orb} = 121.5$ days and a semi-major axis of $a = 0.31$ AU. From \citet{ang14} the habitable zone of \object{Kapteyn's Star} is HZ $\approx 0.126 - 0.23$ AU. Thus, as pointed out by \citet{ang14}, Kapteyn b orbits within the star's HZ and receives a mean instellation of S/S(earth) $\approx 40\%$ of the insolation the Earth receives from the Sun. This is similar to the insolation received by Mars and thus this planet is potentially habitable if it has Greenhouse gas temperature enhancements similar to (or somewhat higher than) Earth ($\sim30$ K). On the other hand, Kapteyn c receives only 12\% of the insolation that the Earth receives and is thus likely too cold for life to evolve, unless its atmosphere supports a large Greenhouse temperature enhancement. Kapteyn b is one of the oldest known (probable) rocky Earth-size exoplanets that is potentially habitable. Moreover, if this discovery is confirmed, the existence of rocky planets hosted by the old, metal-poor \object{Kapteyn's Star} demonstrates that terrestrial planets form in the metal-poor environments of other Halo / Pop II stars over 10 Gyr ago.

However, the results of \citet{ang14} have been questioned by \citet{rob15} who argue against the existence of Kapteyn b. Their analysis of the spectroscopic data indicates possible correlations between stellar rotation and measured radial velocities. They conclude that Kapteyn b is not a planet but an artifact of stellar rotation. More recently, however, \citet{ang15} defended their results and reasserted the existence of this planet.

Recently, \citet{cam15} reported the discovery of the Kepler-444 exoplanet system that consists of five transiting earth and sub-earth size planets hosted by an old, high velocity, metal deficient dwarf K0 star (HIP 94931; LHS 3450). An astereoseismic analysis of the \textit{Kepler} photometry was carried out, yielding an age of $11.2\pm1.0$ Gyr, indicating an Old Disk / Population II host star. Thus, this system and the \object{Kapteyn's Star} system (if confirmed) demonstrate that planet formation in our Galaxy may date back to the earliest eras of star formation -– at times when metal abundances were significantly lower than now. Furthermore, the existence of old exoplanet systems could portent the possibility of complex (and possibly very advanced, even intelligent) life in the solar neighborhood around commonplace stars much older than our Sun. As noted in Section 5, these planets, their atmospheres and possible primordial life, all must survive the expected high levels of XUV radiation when the host stars were young, rotating fast and thus very active.

\section{Ground-based Photometry and Spectroscopy of \protect\object{Kapteyn's Star}: Evidence of Rotation-Induced Activity}

\subsection{Photometry of \protect\object{Kapteyn's Star}}

Kapteyn Star is designated as VZ Pic in the \textit{General Catalog of Variable Stars}\footnotemark\footnotetext{http://www.sai.msu.su/gcvs/cgi-bin/search.htm} where it is classified as a BY Dra-type variable star, whose light variations typically arise from small, rotational brightness variations due to starspots. However, for a star as old (and presumably inactive) as \object{Kapteyn's Star}, only very small light variations are expected.

\begin{figure*}
\epsscale{1}
\plotone{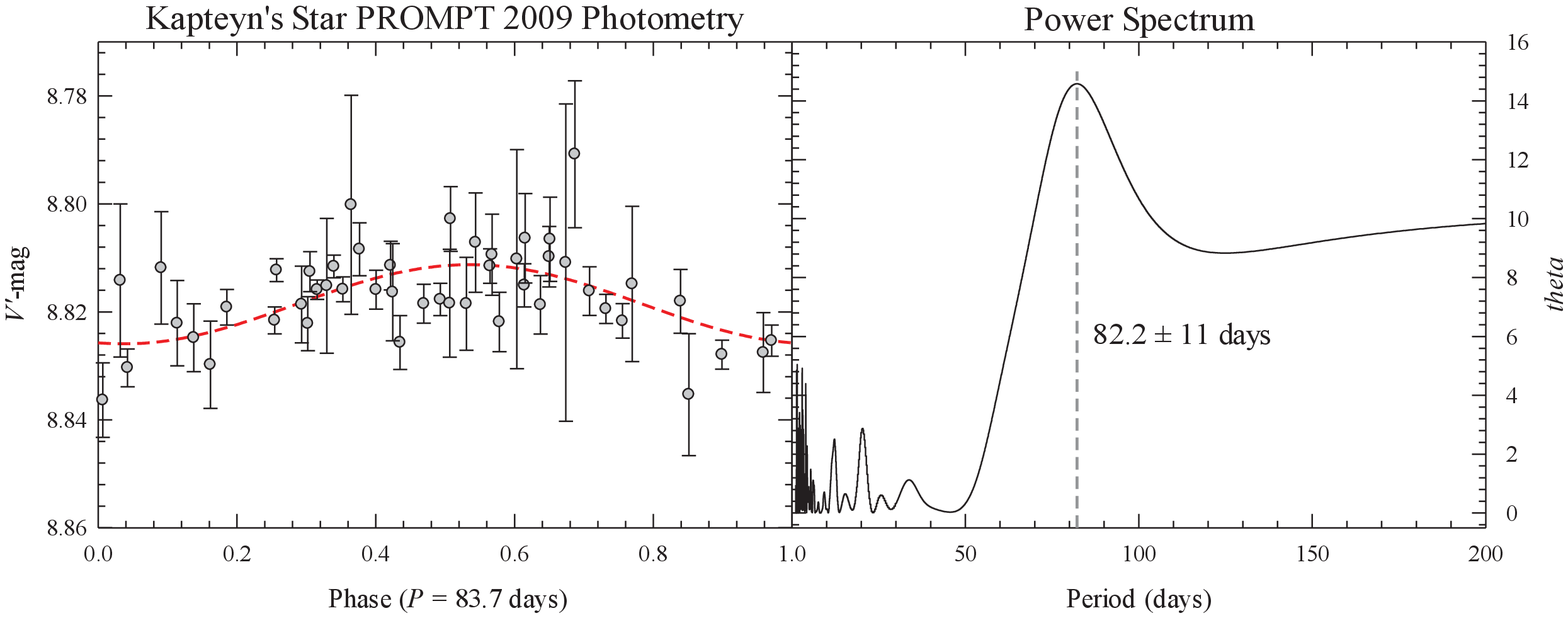}
\caption{\textit{V}-band CCD photometry of \protect\object{Kapteyn's Star} carried out in 2009 with the \textit{PROMPT4} telescope are shown in the left-hand panel. Note that the photometry is phased with the much more precise period of 83.7 days found from analyses of the \textit{HARPS} Ca \textsc{ii} spectra. The right-hand panel shows the power spectrum of the photometry, with the most prominent photometric period of 82.2 days marked. The reasons for the larger uncertainty in the photometric period are the single season of photometry gathered thus far and the small amplitude of the light variations.}
\label{fig1}
\end{figure*}

In 2009 (January through May) we secured high precision time-series CCD photometry of \object{Kapteyn's Star} using the 41-cm \textit{Panchromatic Robotic Optical Monitoring and Polarimetry Telescope} Number 4 (\textit{PROMPT4}; \citealt{rei05}). This telescope is equipped with a 1k$\times$1k Apogee CCD camera \citep{nys09}, resulting in 0.59'' per pixel, for a 10' field of view \citep{rei05}. The star was observed with standard \textit{BVR} filters on $\sim$50 nights over an interval of $\sim$120 days. Each night of observation usually consisted of 4 x 20-second \textit{B}-band, 6 x 8 \textit{V}-band and 9 x 5 \textit{R}-band exposures. Images were reduced in the usual fashion and differential photometry was carried out with \textit{Maxim DL} using 4 nearby reference stars. The \textit{V}-band photometry revealed small, periodic light variations attributed to star spots rotating in and out of view. Thus, this apparently confirms the BY Dra variable star designation. Period searches of these data were conducted using the \textit{Peranso} software's analysis of variance (\textit{ANOVA} -- \citealt{sch96}) statistics algorithm. A quasi-sinusoidal light variation was found with a period of $P_{\rm rot} = 82.2\pm11$ days from the \textit{V}-band data and a light amplitude of $V_{\rm amp} \approx 15\pm4$ mmag. As discussed next in Section 3.2, a very similar but better defined rotation period of $P_{\rm rot}= 83.7\pm0.3$ days is returned from a study of the Ca \textsc{ii} \textit{HK} emissions. The \textit{V}-band data are plotted in Fig.~\ref{fig1}a. While Fig.~\ref{fig1}b shows the resulting period analysis. This photometric period is assumed to be the rotation period of the star. Because of the small light variations and the lack of reliable light curves at different wavelengths, only a rough estimate of the star spot coverage of the star can be made.  Additional spectroscopy and high precision multi-wavelength photometry are needed to secure more precise spot properties. Estimates of the star spot coverage were made using the relations given by \citet{dor87}. Adopting a two spot model, stellar inclination \textit{i} $>$ 60 deg and spot to star temperature ratio of $T_{\rm spot}$/$T_{\rm star}$ $\approx 0.8 - 0.9$, the minimal star spot areal coverage of $\sim1.1 - 3.0$\% was estimated. For comparison, the total fractional spot areas on the Sun are typically $0.1 - 0.2\%$ of the photosphere \citep{cox00}.

It is noteworthy that most of the red dwarf stars in our program (as well as those observed by the \textit{Kepler} Mission) show small, quasi-sinusoidal and periodic light variations presumable also arising from the presence of star spots on their rotating surfaces. Thus, they are also BY Dra variables (e.g. see Engle et al. 2011). In the case of \object{Kapteyn's Star}, the inferred star spots cover more surface area than sunspots do on our Sun, even during solar maxima. Additional high precision, time-series photometry of \object{Kapteyn's Star} is needed to confirm this tentative period, and is currently underway. One problem that can limit photometric precision for \object{Kapteyn's Star} is its large proper motions that cause the star to undergo $\sim$8'' yr$^{-1}$ motions relative to fainter background stars. This can cause photometric measurement errors as the star moves across individual pixels on the CCD. Additionally, there is the possibility of varying fainter background stars being included / excluded in the brightness measures over time.

We also searched for evidence of quasi-periodic light variations in the available online all-sky photometry data sets that include \textit{Hipparcos} \citep{esa97}, and the Warsaw University \textit{All-Sky Automated Survey} (\textit{ASAS}). Because of both the temporal coverage and relatively low photometric precision ($\pm$0.015 mag) of these data, no compelling evidence of significant systematic light variations was found. More recent analysis of additional all-sky photometry from the \textit{Kilo-degree Extremely Little Telescope} (\textit{KELT}) also revealed no coherent, periodic light variations (J. Pepper -- private communication) to the level of typical photometric sensitivity ($\pm0.014$ mag). This was not unexpected, given the large proper motion of \object{Kapteyn's Star}, potential source contaminations and the large pixel scale of the instrument.

\subsection{Rotation Period from Spectroscopy}

Ca \textsc{ii} \textit{HK} (3968.5 and 3933.7\AA) chromospheric emission line measures from \citet{ang14}, obtained with data from the \textit{HARPS} spectrograph, were also searched for variability. These observations were carried out to secure high precision radial velocities of the star to reveal any hosted planets. Ca \textsc{ii} \textit{HK} emissions (commonly given as the \textit{S}-index) have long been recognized as tracers of magnetic activity in cool stars \citep{vau81,wil78}. If active, magnetic (chromospheric) features are asymmetrically distributed over the stellar surface, then periodic variability in the Ca \textsc{ii} emission strength can be used to determine stellar rotation. The \textit{S}-index has been used for decades to study magnetic cycles and rotation rates for cool stars. Studies have shown Ca \textsc{ii} \textit{HK} emission to be a less ambiguous indicator of magnetic activity and variability, displaying a more linear emission response to varying levels of magnetic activity \citep[see][]{mul15,gom14}. The H$\alpha$ (6562.81\AA) feature, on the other hand, is more complicated to interpret since, as magnetic activity increases, the feature goes from being in absorption to eventually emerging into emission above a certain activity threshold \citep[][and references therein]{mul15}. In cool stars, the presence of strong photospheric continuum flux in the H$\alpha$ spectral region adds to the ambiguity in interpreting this feature. As such, we concentrated on analyzing the Ca \textsc{ii} \textit{S}-index for rotation-induced variability, to support the photometric findings, since photometric and \textit{S}-index variations should both arise from differences in stellar active region (starspot, plage) coverage.

\begin{figure*}
\epsscale{1}
\plotone{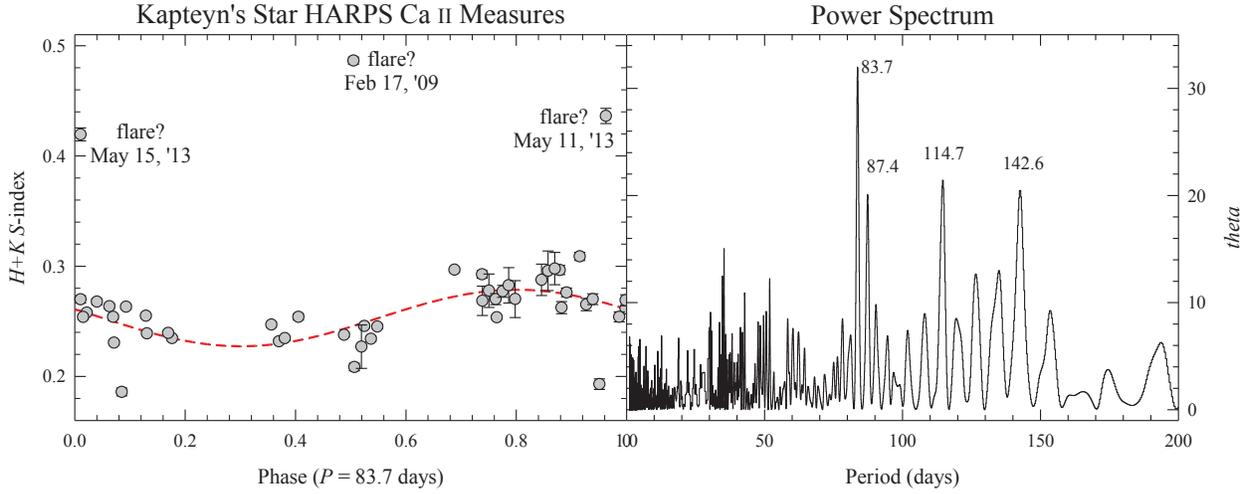}
\caption{The Ca \textsc{ii} \textit{S}-index, derived from \textit{HARPS} spectra as reported by \citet{ang14}, are plotted in the left-hand panel. These data are phased to the most prominent period of 83.7 days returned by \textit{ANOVA} analysis. The power spectrum, along with the four most prominent periods, are shown in the right-hand panel. Potential flare points (marked in the left-hand panel) were excluded from analysis. The period of $\sim$143 days found by \citet{rob15} is one of the prominent periods found, but with a lower probability.}
\label{fig2}
\end{figure*}

Planet hosting stars, in particular, offer excellent opportunities for such studies since they normally possess a wealth of high-quality spectroscopic measures \citep[e.g.][where rotation of the planet-hosting M2V star GJ 176 was also observed in both photometric and Ca \textsc{ii} measures]{for09}. \citet{ang14v} published 95 measures of the Ca \textsc{ii} \textit{S}-index of \object{Kapteyn's Star} taken over the course of a decade, from December 2003 to January 2014. These data were analyzed for periodicites, and the results are given in Fig.~\ref{fig2}. For analysis, three potential flare points were excluded (discussed below), nightly means were determined and, as with the photometry, examined with the \textit{ANOVA} algorithm. A well-defined, periodic variation of $P_{\rm rot} = 83.7\pm0.3$ days was found, in good agreement with the photometric rotation period. Because of the very strong detection, this is the final adopted rotation period. The resulting light curve and power spectrum are given in Fig.~\ref{fig2}. Also, this rotation period agrees with the 71.4 day rotation period we find for GJ 478 from \textit{ASAS-3} data. GJ 478 is an Old Disk / Halo K7V--M0V star \citep{leg92} and, therefore, similar in age and mass to \object{Kapteyn's Star}.

If a rotation period of $P_{\rm rot} = 83.7$ days is adopted, then $v~sin~i \approx 0.2$ km s$^{-1}$, when a stellar radius $R = 0.291\pm0.025 R_{\odot}$ (derived via interferometry by \citet{seg03}; see Table 1) and $i > 60$ deg are assumed. At present, this rotation velocity is too small to be measured spectroscopically. The current practical precision limit to secure the rotation line broadening is $\ga$ 0.5 km s$^{-1}$. For \object{Kapteyn's Star}, only an upper limit of $v~sin~i < 3$ km s$^{-1}$ is available \citep[see][]{ang14}.

As shown in Fig.~\ref{fig2}a, three possible flare events may have occurred: Feb 17, 2009 ($S = 0.487$), May 11, 2013 ($S = 0.436$) and May 15, 2013 ($S = 0.420$). These values are $\sim$1.6 -- 1.8$\times$ the dataset's mean Ca \textsc{ii} \textit{S}-index ($\langle S \rangle \approx 0.25$) for the star. Very rough estimates of flare probability can be made: using the fact that 3 spectra indicated a potential flare, out of the 95 reported, this gives a flare probability of $\sim$3.2\%; using exposure times for the spectra, a flare was observed for 2009 seconds out of the total 60759, thus giving a flare probability of $\sim$3.3\%. At face value, this translates to approximately $3-4$ similar-size flares per day. More intensive observations are needed to confirm flare activity in \object{Kapteyn's Star}. If confirmed, it could indicate that even very old dM stars can continue to generate flares. The flare activity indicated by the Ca \textsc{ii} (\textit{S}-index) data for \object{Kapteyn's Star} appears similar to the flares reported by \citet{kow09,hil10} for older population red dwarfs. Additional Ca \textsc{ii} \textit{HK} or H$\alpha$ spectroscopy and time-series \textit{UBV} photometry (especially \textit{U}-band photometry) are needed to better define the flare activity of \object{Kapteyn's Star}. In the limited data sets, no super-flares (i.e. flares with total energies E $>$ 10$^{\rm 33}$ ergs) have been observed. 


\section{Coronal X-ray Emissions -- Still Active, after all these years}

\object{Kapteyn's Star} was discovered to be a weak, soft X-ray source from observations carried out with \textit{ROSAT} during the 1990s. X-ray luminosities of  $L_{\rm X} = 2.4\times10^{26}$ erg s$^{-1}$ to $5.6\times10^{26}$ erg s$^{-1}$ were determined \citep{mic97,sch04} from the observations. Although not detected by the \textit{Einstein} Observatory (observations carried out during 1979), \citealt{mic97} give an upper limit of $L_{\rm X} < 3.2\times10^{26}$ erg s$^{-1}$. These X-ray luminosities are similar to those of other old (inactive) red dwarfs (see Engle and Guinan 2011). The $L_{\rm X}$ values are also similar to the Sun near its activity minimum (the Sun has a full range of $4\times10^{26} < L_{\rm X} < 2.0\times10^{27}$ ergs s$^{-1}$ from \citealt{ayr14}).

As part of the \textit{Living with a Red Dwarf} Program, new X-ray observations were carried out of \object{Kapteyn's Star} (along with several other red dwarfs) with the \textit{Chandra X-ray Observatory}. The X-ray observation of \object{Kapteyn's Star} was made on 14 October, 2012, for 5.0 ksec using the \textit{ACIS-I} instrument. Since the data were from the single, rather short exposure of a nearby, well-centered object, no special processing was required and the pipeline-processed \textit{Chandra} data products were used for analysis. Extraction and grouping of the spectral products was carried out in the usual fashion, with \textit{CIAO} v4.7 and the \texttt{specextract} routine. A small amount of ambiguity exists when analyzing the data, because of the softness of the source, the short exposure and resulting low number of counts. As such, two separate models were fit to the energy distribution using \textit{Sherpa}: two simultaneous \texttt{xsmekal}\footnotemark\footnotetext{http://cxc.harvard.edu/sherpa/ahelp/xsmekal.html} models were fit to data grouped with 5 energy channels per bin (model 2T5 in Table 2), and then also fit to data grouped with 10 energy channels per bin (model 2T10, also in Table 2).

\begin{figure*}
\epsscale{1}
\plotone{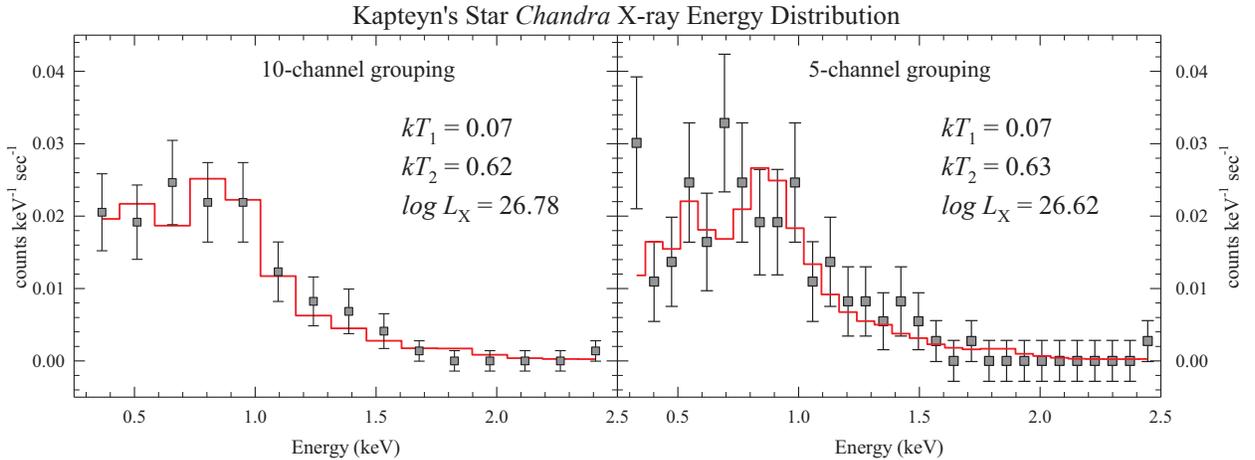}
\caption{Data from the \textit{Chandra ACIS-I} observation of \protect\object{Kapteyn's Star} are show with two separate groupings: 10-channel grouping in the left-hand panel and 5-channel grouping in the right. The best-fitting, two-temperature \textit{Sherpa} \texttt{xsmekal} models are plotted (red lines) with each dataset, and the relevant parameters are given in the plots (also see Table~\ref{tbl2}).}
\label{fig3}
\end{figure*}

A neutral hydrogen ($N_{\rm H}$) value of $8\times10^{17}$ cm$^{-2}$ was adopted, based upon the distance to \object{Kapteyn's Star} and the data provided by \citet{par84}, for both models. The data were best fit by a two-temperature model of $kT_{\rm 1} = 0.07$ and $kT_{\rm 2} = 0.62$ keV. The X-ray flux was determined to be f$_{\rm X} = 2.8\times10^{-13}$ erg s$^{-1}$ cm$^{-2}$ which, using the distance to \object{Kapteyn's Star} of $d = 3.91$ pc, yields $L_{\rm X} = 5.13\times10^{26}$ erg s$^{-1}$ ($log~L_{\rm X} = 26.77$). The plots of count rate vs. energy and fits to these data are given in Fig.~\ref{fig3}.

Recently, \citet{joh15} determined a tight correlation between X-ray surface flux $F_{\rm X}$ and coronal temperature $T_{\rm cor}$ for a sample of G,K,M stars. The relation they find is: $T_{\rm cor} = 0.11(F_{\rm X})^{0.26}$ in which $F_{\rm X}$ is in c.g.s. units. Using this relation, the mean $L_{\rm X}$ and the stellar radius of \object{Kapteyn's Star} given in Table 1, returns: $\langle F_{\rm X} \rangle = 9.72\times10^4$ erg s$^{-1}$ cm$^{-2}$. Applying this relation to \object{Kapteyn's Star} yields a mean coronal temperature of $\langle T_{\rm cor} \rangle = 2.2$ MK. This is in generally good agreement with the weighted coronal temperatures indicated by our modeling of the \textit{Chandra} data of $kT_{\rm 1} = 0.07$ keV ($= 0.81$ MK) and $kT_{\rm 2} = 0.62$ keV ($= 7.8$ MK). Also included in this plot are X-ray irradiances given for a reference distance of 1 AU ($f_{\rm X}~(1 AU))$. As discussed earlier, the near-mid HZ for \object{Kapteyn's Star} is $a=0.17$ AU. At this distance, by applying the inverse square law, the X-ray irradiance would be 35$\times$ larger than at 1 AU.

Magnetic activity cycles (and flares) are common in cool stars, so the observed variations among the three X-ray measures are understood, although there is clearly not enough data to begin analyzing a possible cycle. Taking the two \textit{ROSAT} measures and the averaged \textit{Chandra} results, the mean X-ray activity level of \object{Kapteyn's Star} is: $\langle {\rm f}_{\rm X} \rangle = 2.40\pm0.77\times10^{-13}$ erg s$^{-1}$ cm$^{-2}$; $\langle L_{\rm X} \rangle = 4.40\pm1.41\times10^{26}$ erg s$^{-1}$; $\langle log~L_{\rm X} \rangle = 26.64\pm0.16$. The ratio of $\langle L_{\rm X} \rangle$ to the total bolometric luminosity of \object{Kapteyn's Star} is $\langle L_{\rm X} \rangle / L_{\rm bol} = 9.53\times10^{-6}$. For comparison, and adopting a mean solar X-ray luminosity of $\langle L_{\rm X} \rangle$ = $1.2\times10^{27}$ erg s$^{-1}$ \citet{ayr14}, the $\langle L_{\rm X} / L_{\rm bol} \rangle$ of the Sun $= 3.13\times10^{-7}$. Thus, \object{Kapteyn's Star} has a $\langle L_{\rm X} \rangle / L_{\rm bol} \approx 30\times$ larger than that of the mean Sun.

\begin{figure*}
\epsscale{1}
\plotone{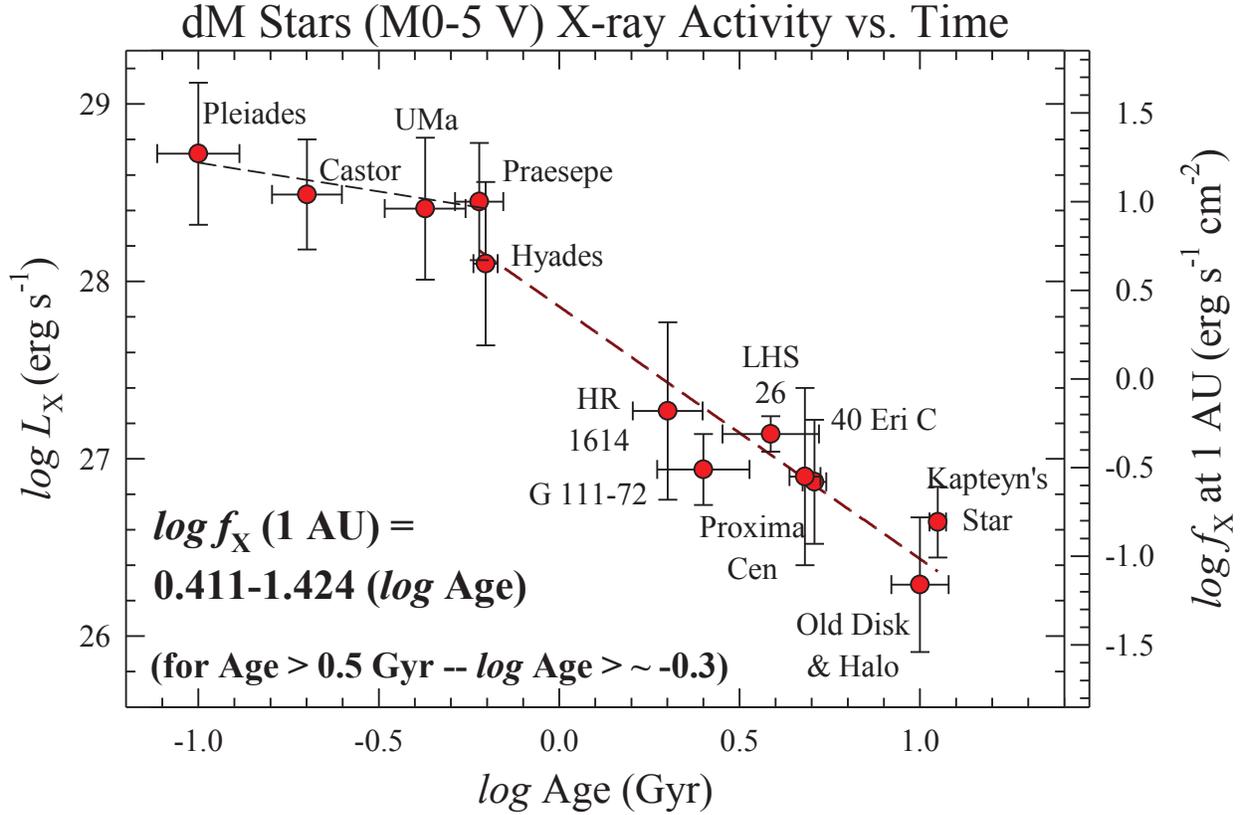}
\caption{The decline in dM0--5 star X-ray activity over time is plotted. The y-axis is given in terms of both $log~L_{\rm X}$, and the $log~f_{\rm X}$ at a reference distance of 1 AU. As discussed by \citet{eng11}, the ages of red dwarf stars are determined from memberships in clusters, moving groups, and wide binaries and common proper motions pairs in which the companions have well-determined isochronal or astereoseismic ages. Several of the stars plotted (e.g. 40 Eri C, GJ 176, LHS 26, G111-72) have reliable ages from memberships in wide binaries where white dwarf components have ages estimated from their main-sequence lifetimes and cooling times\citep{gar11,zha11}. The position of \protect\object{Kapteyn's Star} is labeled in the plot. A least-squares fit to these data has been made (shown by the line in the plot) between X-ray emission and age: $log~L_{\rm X} = 27.857-1.424(log~Age[Gyr])$ or alternatively $log~f_{\rm X}$ (1 AU) = $0.411-1.424(log~{\rm Age[Gyr]})$. As shown, the X-ray emission levels decrease by $\sim71\times$ over a nominal age range of 0.5--10.0 Gyr. This plot gives a useful relation for determining X-ray irradiance vs. age for M0--5 V stars. A subsequent paper is in preparation to discuss the details of this relation as well as providing a calibration of rotation period with stellar age. For reference purposes, converting from $log~L_{\rm X}$ to $f_{\rm X}$ (1 AU) is done via $f_{\rm X}~{\rm(1 AU)} = 10^{log~L_{\rm X}} / 2.812\times10^{\rm 27}$. To adjust $f_{\rm X}$ (1.0 AU) to the near mid-HZ of \protect\object{Kapteyn's Star} (0.17 AU), a 34.6$\times$ increase is applied from the inverse square law.}
\label{fig4}
\end{figure*}

In Fig.~\ref{fig4}, we combined our measures of the X-ray luminosity of \object{Kapteyn's Star} with other red dwarfs in our \textit{Living with a Red Dwarf} program that have reliable ages. As discussed by \citet{eng11}, the ages of red dwarf program stars are determined from memberships in clusters, moving groups, as well as in wide binaries and common proper motions pairs where the companions have well-determined isochronal or astereoseismic ages. Several of the stars plotted in Fig.~\ref{fig4} (e.g. 40 Eri C, GJ 176, LHS 26, G111-72) have reliable ages from memberships in wide binaries with white dwarf components, which have ages determined from their main-sequence lifetimes and cooling times \citep{gar11,zha11}. For the oldest stars (like \object{Kapteyn's Star}) space motions and (low) metalicity are also utilized to infer ages. As shown in the figure, the X-ray flux of red dwarfs (i.e. M0--5 V stars) decreases sharply with increasing age and longer rotation periods. For example, from the least squares fit to these data (as given in Fig.~\ref{fig4}), the X-ray flux (X-ray luminosity)  decreases by $\sim$71$\times$ from a young reference age of $\sim$0.5 Gyr (and estimated minimum level of saturated\footnotemark\footnotetext{Saturation is the age range where coronal activity will no longer increase with decreasing age and faster rotation, as the coronal volume is already filled with magnetic structures. For red dwarfs, this occurs at Ages below $\sim$0.5--0.8 Gyr \citep{kir07}.} activity, $log~L_{\rm X}\approx28.3$) to the an old reference age of $\sim$10 Gyr. This result is in good accord with the results of the earlier activity-rotation-age study of M dwarfs by \citet{kir07}. This study uses photometric data from the ASAS program and archival X-ray, measures mostly from \textit{ROSAT}. They find that the mean X-ray flux of old-disk ($\sim$10 Gyr) dM stars is about $1.7\pm0.5$ dex lower than young disk stars (ages $<$ 1 Gyr) in their sample. Unfortunately, the age estimates of the stars are based on space motions and are somewhat uncertain. This relationship gives a decrease in L$_{\rm X}$ of $\sim$50 times over an estimated age spread of $\sim$10 Gyr.

\section{\textit{HST-COS} Spectra of \protect\object{Kapteyn's Star}: FUV--UV Chromospheric \& Transition-Region Line Emissions and Diagnostics}

During \textit{HST} Cycle 20 we carried out \textit{COS} FUV--UV (1150 -- 3100\AA) spectrophotometry of a carefully selected sample of M1--5 V stars with well-determined ages. The goals were to study the evolution of their dynamo-generated UV emissions with age, and to better understand the heating and energetics of their atmospheres. These data also allow studies of UV radiation and its effects on the environments of extrasolar planets hosted by red dwarfs, and on the possible origin and evolution of life on such planets. This program complemented \textit{Chandra} observations of the same stars to secure coronal X-ray fluxes. \object{Kapteyn's Star} is an excellent target for such a program since it is a nearby, important proxy for Pop II metal poor stars with a large radial velocity of +245 km/sec. At this velocity the Ly$\alpha$ line will be red-shifted by $\sim$1\AA, moving most of the stellar emission line away from the ISM and geo-coronal features. This allows the stellar Ly$\alpha$ flux to be unambiguously measured. Also, the Mg \textsc{ii} \textit{hk} 2795.5 and 2802.7\AA~emission features are Doppler-shifted by $\sim2.3$\AA~ -- well clear of any significant ISM absorption. 

\begin{figure*}
\epsscale{1}
\plotone{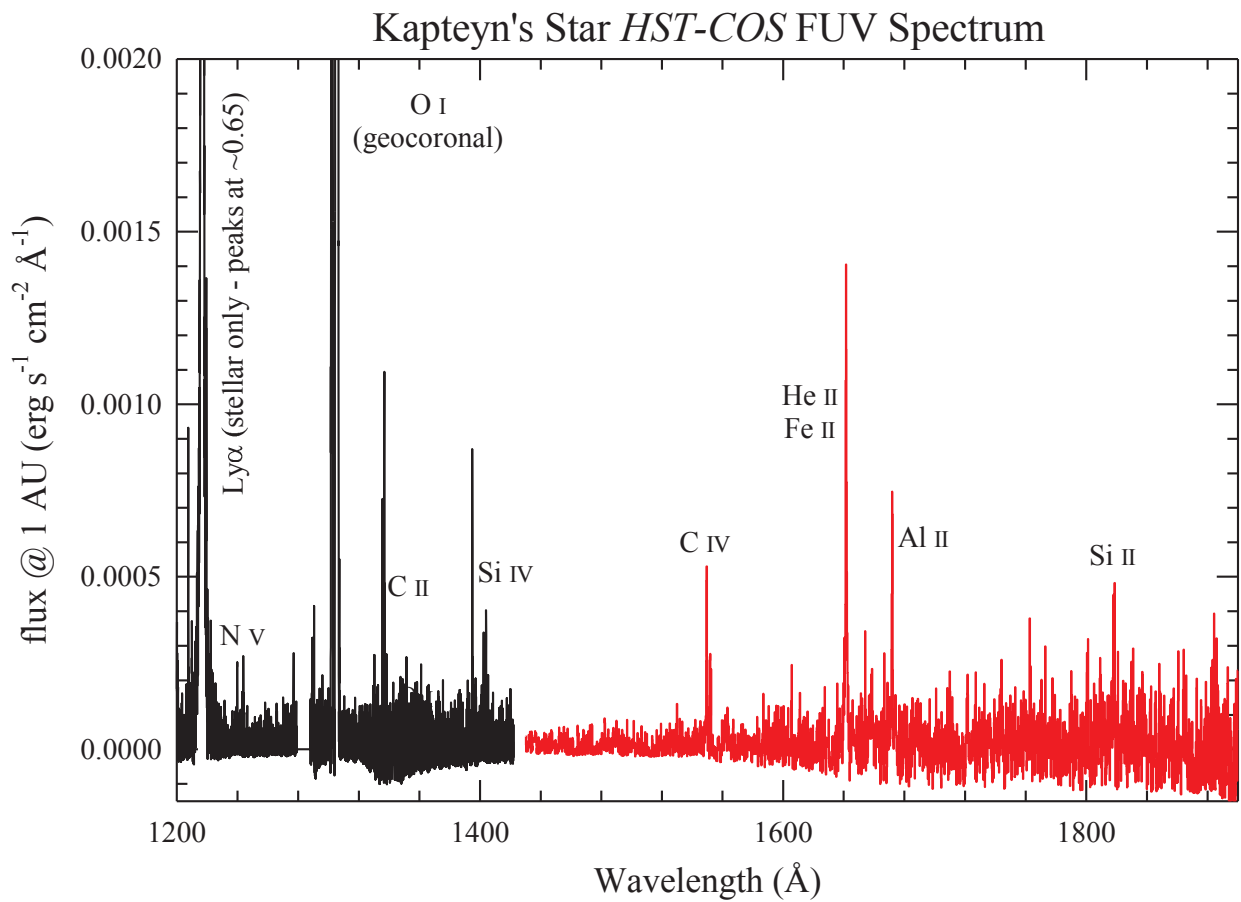}
\caption{The \textit{HST-COS} FUV (G140L \& G130M) spectrum of \protect\object{Kapteyn's Star} obtained as part of this program. As indicated in the plot, the geo-coronal Ly$\alpha$ emission feature has been removed, leaving only the stellar emissions. Some of the most prominent chromospheric and transition region emission lines (Ly$\alpha$ included) have been labeled. The entire wavelength region plotted is free of photospheric continuum flux.}
\label{fig5}
\end{figure*}

\begin{figure*}
\epsscale{1}
\plotone{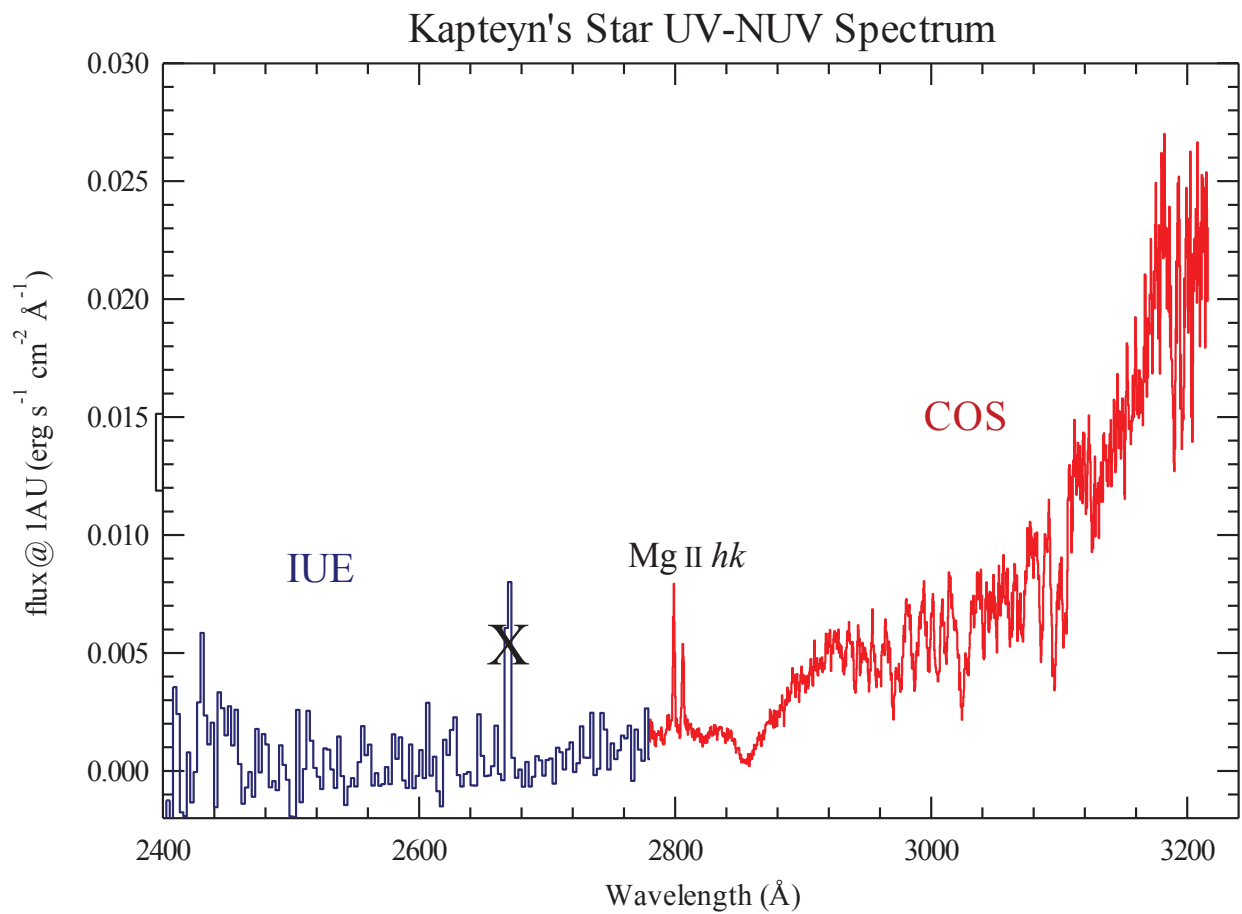}
\caption{The merged \textit{HST-COS} and \textit{IUE} NUV spectrum of \protect\object{Kapteyn's Star}. The ``feature'' visible near $\sim$2680\AA~is likely caused by a cosmic ray hit, so it has been marked with an X. As seen in the plot, the photosphere begins contributing continum flux longward of $\sim$2700\AA. The Mg \textsc{ii} lines, the most prominent emission feature in the NUV region, are marked. }
\label{fig6}
\end{figure*}

The wavelength region of \textit{COS} G130M \& G140L/230L include many important diagnostic chromospheric \& Transition-Region (TR) emission lines, spanning plasma temperatures from $\sim$8,000 -- 300,000 K (e.g. C \textsc{iii} 1175\AA~triplet, Si \textsc{iii} 1206, H \textsc{i} Ly$\alpha$ 1215.67, N \textsc{v} 1240, Si \textsc{ii} 1295--1306 (fluorescence by O \textsc{i}), O \textsc{i} 1302--1306, Si \textsc{ii} 1309, C \textsc{ii} 1335,  Si \textsc{iv} 1400 doublet, Si \textsc{ii} 1526, C \textsc{iv} 1550 doublet, He \textsc{ii} 1640, and C \textsc{i} 1660). The full analysis of these temperature-dependent emission lines will be included with the analysis of other program stars in a subsequent paper. The NUV ($\sim$2000 -- 3200\AA) elements primarily are employed to measure the strong chromospheric Mg \textsc{ii} \textit{hk} emission and the stellar continuum longward of 2600\AA. The Mg \textsc{ii} \textit{hk} emission is, like Ly$\alpha$, an important chromosphere cooling channel. Figures~\ref{fig5} and~\ref{fig6} show plots of the FUV and NUV spectrophotometry of the star. Table 3 gives the integrated fluxes of the strongest emission features. 

The \textit{COS} G130M and G140L spectra are shown in Fig.~\ref{fig5}. The flux received at a stellar reference distance of 1 AU (irradiance) is plotted and the plot covers most of the FUV region from $\sim$1200 -- 1900\AA. The strongest emission features are identified. It should be noted that the plotted Ly$\alpha$ emission feature is from the star and excludes Ly$\alpha$ geo-coronal emission and ISM absorption (the determination of the Ly$\alpha$ emission of the star is discussed in the next section). Fig.~\ref{fig6} shows the NUV spectrum of \object{Kapteyn's Star}. The $COS$ 230L spectrum is plotted and combined with a low dispersion, large aperture \textit{International Ultraviolet Explorer} (\textit{IUE}) LWP spectrum (LWP 11429L) available in the \textit{MAST-IUE} archive. Again, the stellar flux at 1 AU (irradiance) is plotted. The Mg \textsc{ii} \textit{hk} emission feature is identified. As shown in the plot, like other red dwarf stars, \object{Kapteyn's Star} does not have significant NUV flux until reaching wavelengths longer than $\sim2700$\AA. This is much different than solar-type stars like the Sun that have strong NUV photospheric continuum fluxes. The low NUV fluxes of red dwarfs (including \object{Kapteyn's Star}) compared to Sun and solar-type stars will have consequences in driving photochemical reactions, including ozone production, in the atmospheres of hosted HZ planets.

\subsection{Chromospheric H {\mdseries\textsc{i}} Ly$\alpha$ Emission}

A third \textit{HST} orbit on \object{Kapteyn's Star} was employed to observe the Ly$\alpha$ region with the \textit{COS} G130M spectral element. This observation gives the resolution necessary to perform a detailed study of the Ly$\alpha$ feature, comparing it to the reconstructed Ly$\alpha$ fluxes modeled by (\citet{woo05,fra13} and references therein) and to test acoustic heating models for stars with very slow rotations. This observation provides the first ``clean'' integrated Ly$\alpha$ flux for a star (other than the Sun).

The chromospheric H \textsc{i} Ly$\alpha$ emission line is by far the strongest emission feature in the solar EUV/FUV spectrum \citep[see][and references therein]{tia13}. This line is also dominant in the EUV/FUV of cool (G, K, M) main-sequence stars \citep[see][]{lin14}. Because of its importance to solar-chromospheric physics and also for heating and ionizing effects on Earth's upper atmosphere and ionosphere, the solar Ly$\alpha$ emission has been studied for over 60 years with rockets, satellites, and solar missions \citep[e.g.][and references therein]{tia13}. As such, Ly$\alpha$ emission is the most important contributor to the heating, ionization and photochemistry of the upper atmospheres of planets. In particular the Ly$\alpha$ flux plays a major role in the photodissociation of important molecules such as H$_{\rm 2}$O, CO$_{\rm 2}$ and CH$_{\rm 4}$. For the Sun, the Ly$\alpha$ emission line contributes the overwhelming majority of the EUV/FUV 100 -- 1250\AA~ flux. Though the H \textsc{i} Ly$\beta$ (1028\AA) is strong in the Sun, and it can easily be measured without interference from H \textsc{i} ISM absorption, it is still significantly weaker than Ly$\alpha$ (solar Ly$\alpha$ /Ly$\beta$ ratio $\approx 212 - 230$ from \textit{SOHO-SUMER} -- \citealt{tia13}). However, due to the very strong H \textsc{i} local ISM absorption and possible contamination from geo-coronal emission, the Ly$\alpha$ line is very difficult to measure even for nearby stars. Nonetheless, over the last decade, measures of Ly$\alpha$ emission fluxes have been obtained with \textit{HST} \citep[e.g.][]{woo05,fra13,lin14}. These stellar Ly$\alpha$ integrated fluxes have been reconstructed from \textit{HST} \textit{STIS} and \textit{COS} spectra for a small sample of mainly main-sequence G, K \& M stars. The stellar Ly$\alpha$ emission is calculated by taking into account the ISM absorption and removing the sharp geo-coronal emission contributions. For the cases in which the stellar Ly$\alpha$ emission is weak (e.g. older, inactive stars), the corrections to the observed Ly$\alpha$ profile can be very large. The model-reconstructed stellar fluxes can be more than $10-20\times$ the observed, uncorrected profile. The applied corrections lead to uncertainties of 10--30\% in the resulting, integrated Ly$\alpha$ emission fluxes.

\begin{figure*}
\epsscale{1}
\plotone{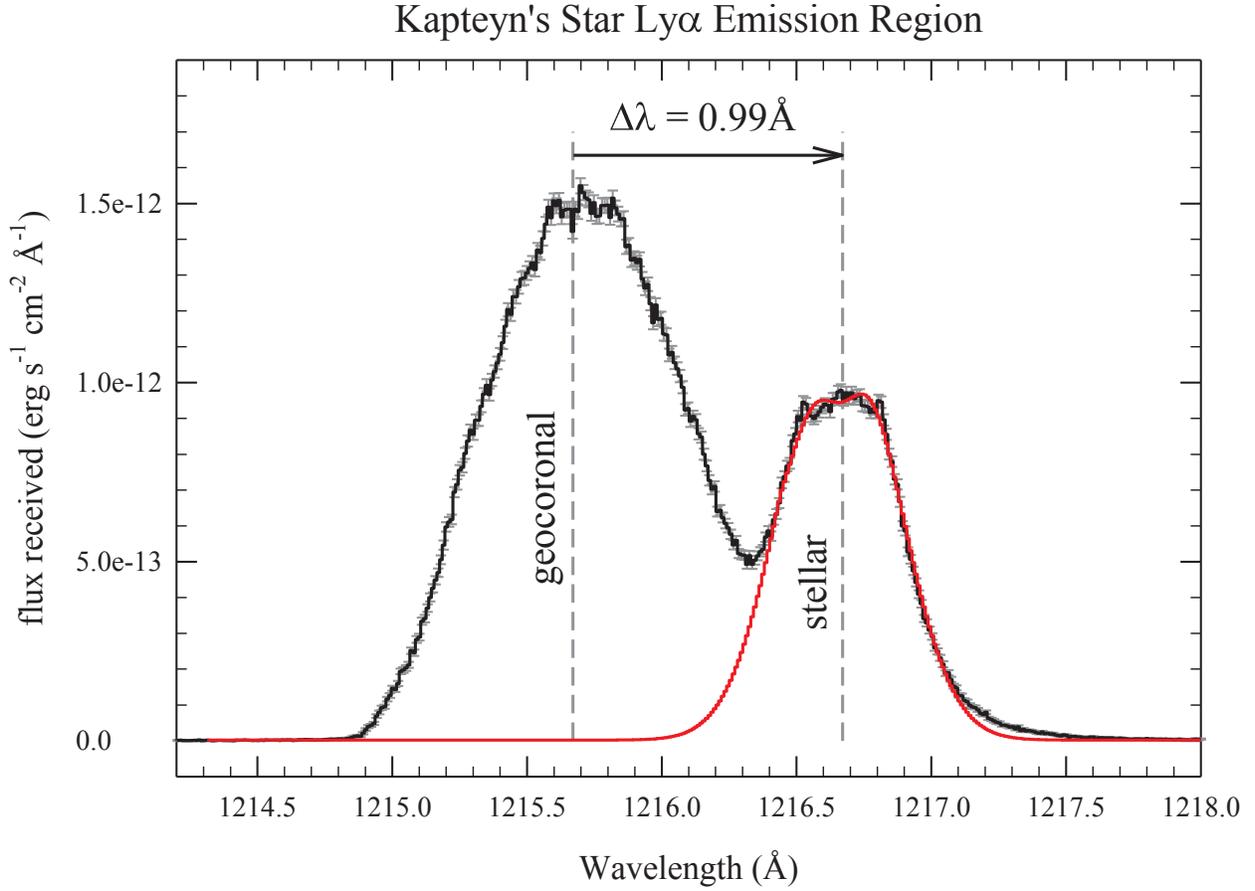}
\caption{A close-up of the Ly$\alpha$ region in the \textit{HST-COS} G130M spectrum, showing both the stellar and geo-coronal emission features. As seen in the plot, and discussed in the text, the high RV of \protect\object{Kapteyn's Star} results in a Doppler shift of $\sim$0.99\AA. The ``stellar only'' emission profile is plotted in red. The integrated flux of the Ly$\alpha$ emission feature, adjusted to a reference stellar distance of 1 AU -- $f_{\rm Ly\alpha}$ (1 AU) = 0.347 erg s$^{-1}$ cm$^{2}$ -- is given in Table~\ref{tbl1}.}
\label{fig7}
\end{figure*}

Because of the high radial velocity of \object{Kapteyn's Star} ($RV = +245.19$ km s$^{-1}$ -- \citealt{nid02}), the stellar Ly$\alpha$ emission feature is red shifted by $\sim1$\AA. This redshift displaces the stellar emission feature from both the ISM absorption and geo-coronal emission features (see Fig.~\ref{fig7}), allowing a much more `clean' determination of the stellar activity. \object{Kapteyn's Star} was observed with \textit{COS} (G130M) to produce a spectrum of sufficient resolution to model the Ly$\alpha$ feature. Still, care had to be taken with the modeling, since there is overlap between the red wing of the geo-coronal feature and the blue wing of the stellar feature (Fig.~\ref{fig7}). To account for this overlap, a multi-component Gaussian fit was run to both features simultaneously. This permitted accurate fitting of the non-Gaussian shape of the geo-coronal feature, in addition to the broad peak and wings of the stellar feature.

\object{Kapteyn's Star} is $\sim$2.5$\times$ the age of the Sun and rotates more than 3$\times$ slower. Nonetheless, its coronal X-ray and chromospheric FUV Ly$\alpha$ surface fluxes are comparable to the Sun near its activity minimum. The ratio of the X-ray and Ly$\alpha$ surface fluxes of \object{Kapteyn's Star} relative to the mean Sun are $F_{\rm KS}$ / $F_{\rm Sun}$ = 4.33 and 0.59, respectively. In spite of its slow rotation, this magnetic-related activity most likely arises from the star's deeper convective zone and more efficient dynamo.

The (liquid water) habitable zone for \object{Kapteyn's Star} is HZ $\approx 0.13 - 0.24$ AU \citep{ang14}. For purposes of comparison, we adopt the near mid-HZ of 0.17 AU, chosen to also be close to the semi-major axis ($a = 0.168$ AU) of the exoplanet candidate Kapteyn b. The mean X-ray (0.3 -- 2.5 keV) irradiance at 0.17 AU is $f_{\rm X}~{\rm (0.17 AU)}=5.41$ erg s$^{-1}$ cm$^{-2}$, while the mean X-ray irradiance of the Earth from the Sun is 1.32 erg s$^{-1}$ cm$^{-2}$. Thus, even at the very old age of \object{Kapteyn's Star}, a planet near the mid-HZ receives, on average, an X-ray flux $\sim$4$\times$ greater than the mean level that the Earth receives from the Sun. The FUV irradiance, defined by the dominant Ly$\alpha$ feature, near the mid-HZ (0.17 AU) of \object{Kapteyn's Star} is $f_{\rm Ly\alpha}~{\rm (0.17 AU)} = 12.0$ erg s$^{-1}$ cm$^{-2}$; for the Sun-Earth, the mean Ly$\alpha$ irradiance is $\sim$7.00 erg s$^{-1}$ cm$^{-2}$. Thus, a planet near the mid-HZ (e.g. Kapteyn b) receives (on average) $\sim1.7\times$ more Ly$\alpha$ FUV radiation than the corresponding average solar Ly$\alpha$ irradiance of the Earth. 

The XUV irradiance near the mid-HZ of \object{Kapteyn's Star} is higher on average than the Earth receives from the present Sun, but not extreme. Thus, these X-ray and Ly$\alpha$ irradiances do not appear to be limiting factors for the survival of planetary atmospheres (or life) on a magnetically protected planet like the Earth, located near the mid-HZ of \object{Kapteyn's Star} at its present (old) age. But this was not always the case. \object{Kapteyn's Star}, when young (age $<$ 0.5 Gyr) is expected to have rotated fast  ($P_{\rm rot} < 10$ days) and correspondingly to have been more magnetically-active than today (see Fig.~\ref{fig4} and ~\ref{fig8}). The X-ray and FUV Ly$\alpha$ fluxes of the young \object{Kapteyn's Star} can be estimated from other young ($\sim0.1$ Gyr $<$ ages $<$ 0.5 Gyr) red dwarfs -- such as AD Leo and EV Lac. The mean values of $f_{\rm X}$ (1.0 AU) = 19.3 erg s$^{-1}$ cm$^{-2}$ and $f_{\rm Ly\alpha}$ (1.0 AU) = 6.2 erg s$^{-1}$ cm$^{-2}$, respectively, are obtained for AD Leo and EV Lac from \citet{lin14}. Using these stars as young proxies, the corresponding mean X-ray and FUV irradiances at $\sim$0.17 AU (the location of Kapteyn b) are estimated to be $\sim$123$\times$ and $\sim$18$\times$ higher, respectively, in the past when \object{Kapteyn's Star} was $\sim$11 Gyr younger. 

These high XUV HZ irradiances inferred for \object{Kapteyn's Star} during its youth, as well as for other young red dwarfs, appear to be unfavorable for the development of life on a hosted, inner-HZ planet unless the planet is protected by a significant geo-magnetosphere. Such high stellar XUV fluxes (in addition to the possible large, frequent flares) ionize and possibly erode a HZ planet's atmosphere even before life has time to develop \citep[e.g.][and references therein]{lam08,eng11}.

From a number of studies of flare characteristics for young stars, based primarily on \textit{Kepler} Mission data, it appears that young G,K,M stars show frequent, energetic flares (e.g. see Walkowicz et al. 2011; Shibayama et al. 2013; Candelaresi et al. 2014; Maehara et al. 2015; Guinan et al. 2015 and references therein). Moreover, the flare rates of red dwarfs indicate that flares persist to older ages, when compared to G-type stars. Flares -- and the resulting strong XUV radiation, plasma (proton) ejection events, and possible associated coronal mass ejections (CMEs) -- could have deleterious effects on the atmospheres and even the ground-level environments of close-in HZ planets. The possible long-term persistence of strong flaring and associated CME events for dM stars and their effects on hosted HZ planets are the main concerns regarding the development and evolution of life on such worlds.

On the other hand, \citet{seg10} studied the effects of a strong stellar flare on the atmosphere of an Earth-like planet orbiting within the HZ of a young, very active dM3 star. They used a well observed, large flare on AD Leo (one of the most active dM (flare) stars) to study the XUV effects of the flare, including its high energy protons, on the atmosphere and atmospheric chemistry of the impacted planet. They conclude that such flares would not significantly affect the potential habitability of the planet or be a direct hazard to life on the surface. They carried out the modeling on a planet identical to Earth, but without a magnetic field. Although such flares can affect a HZ planet's atmosphere and ozone temporarily, these effects appear to be short-lived and not harmful to most surface life.

Historically, older red dwarfs have not yet been extensively studied for flare activity, primarily because they appear ``dull'' with rather weak emission lines and also their ages are difficult to determine reliably. A notable exception is the thorough analysis of Sloan Digital Sky Survey photometry by \citet{kow09}, where flare rates for M dwarfs were found to decrease by an order of magnitude or more as distance from the galactic plane (taken as a proxy for stellar age) increased. Additionally, the solar-age red dwarf Proxima Centauri (age $\sim 5.1\pm0.5$ Gyr -- from membership in the $\alpha$ Cen system) has been extensively studied for flares in the X-ray--UV regions \citep[][and references therein]{fuh11}. These analyses indicate a significant X-ray--UV flare rate \citep[][and references therein]{hai83,wal81}. The examination of the available \textit{VR} photometry of \object{Kapteyn's Star} shows no significant flares in the optical region. But the overall on-star sample-time is relatively small, and no high-precision, time-series flare searches have been carried out in either the \textit{U} or \textit{B} bands where flares are easier to detect since they emit a higher flux at these wavelengths, while the stellar continuum is much weaker. However, there is some evidence for possible ``chromosphere flares'' from the Ca \textsc{ii} spectroscopic observations discussed earlier.

However, on a positive note for potential habitability, these inferred high XUV HZ irradiances for the young \object{Kapteyn's Star} (and for similar young red dwarfs) are about the same as endured by the Earth some $\sim4.0 - 4.5$ Gyr ago when the Sun was young and very magnetically active. As shown from \citet{rib05}, the inferred X-ray and FUV irradiances received by the young Earth were $\sim200 - 300\times$ and $\sim10 - 20\times$ higher, respectively, than now \citep{gui03,rib05}. On the other hand, young dM stars such as AD Leo and EV Lac have frequent large flares (AD Leo: \citealt{ost08,hun12} -- for EV Lac: \citealt{ost05,ost10}) that can cause considerable damage and loss to the atmospheres (and life) on close-in HZ planets such as Kapteyn b when its host was young. But young solar-type stars (age $<$ 300 Myr) also appear to have large (super)flares. By extension, the young Sun is expected to have also generated superflares with total energies of $E_{\rm total} > 10^{33}$ erg \citep[see][and references therein]{ayr15,not13}. In all cases it appears necessary that the HZ planet sustain a strong geomagnetic field and thick atmosphere. This will help shield it from the deleterious effects of the host star's high XUV radiation, strong stellar winds and frequent flares / coronal mass ejections, which are especially strong when the host star is young and very active. This is an even more important factor for planets hosted by dM stars, since their HZs are close ($<0.2$ AU) and their periods of strong stellar activity may persist for longer times.

\subsection{Evidence of Flares and/or Activity Cycles}

The observed variations of the star in coronal X-ray emission (from $L_{\rm X} = 2.4 - 6.0\times10^{26}$ erg s$^{-1}$) and the possible flare-like variations in the Ca \textsc{ii} \textit{HK} emission (from an average value of $\langle S \rangle = 0.25$ up to $\langle S \rangle \approx 0.43 - 0.48$ on three individual observations) indicate that the star is still active. To search for corroborative evidence of variability, we compared our chromospheric Mg \textsc{ii} \textit{hk} emission fluxes with a recently available \textit{HST-STIS} spectrum containing this feature. A high resolution spectrum of \object{Kapteyn's Star} was secured with the \textit{STIS} E230H element on Aug. 18, 2014 -- nearly a year after our \textit{COS} observations. This snapshot spectrum  (PI: Redfield; Program ID: 13332) is centered on the Mg \textsc{ii} \textit{hk} emission feature. We measured an integrated emission flux (the sum of both lines) of $f_{\rm Mg II} = 7.55\times10^{-14}$ erg s$^{-1}$ cm$^{-2}$ \AA$^{-1}$ from this \textit{STIS} spectrum. Surprisingly, this flux is over twice the Mg \textsc{ii} flux ($= 3.07\times10^{-14}$ erg s$^{-1}$ cm$^{-2}$ \AA$^{-1}$) returned from our \textit{COS} G230L spectrum on Sept. 22, 2013. This relatively large variation could arise from a flare or may be part of a long-term activity cycle. The relative observed change in the Mg \textsc{ii} fluxes are similar in magnitude to the three possible flare events seen in the Ca \textsc{ii} lines. 

\subsection{Variations of Ly$\alpha$ Flux with Stellar Age: Tracing the Evolution of Red Dwarf FUV Emissions with Age}

We analyzed the FUV Ly$\alpha$ emission flux levels (vs. stellar age) for dM stars. Including \object{Kapteyn's Star}, only 10 dM stars have Ly$\alpha$ flux measures. We used the measures of the Ly$\alpha$ flux at 1 AU given by \citet{lin14}, and combined these with our determination of the Ly$\alpha$ flux for \object{Kapteyn's Star}. We did not include AU Mic in this study because it is a very young, probable pre-main-sequence, member of the $\beta$ Pic moving group \citep{bar99} with an estimated age of $\sim23\pm3$ Myr \citep{mam14}. We estimated the ages for the other stars using various criteria such as membership in Moving Groups along with rapid rotation (AD Leo \& EV Lac), membership in a wide multiple star system of known age (Proxima Cen), and large space motions/low metals (\object{Kapteyn's Star}). The ages of the other stars were inferred from rotation-age relations or (less certain) activity-age relations. The results are given in Table 4. The Mg \textsc{ii} \textit{hk} fluxes are also given, as they are the most prominent emission feature in the NUV. The integrated flux measures are computed for 1 AU (as a reference distance) and also for 0.17 AU, near the mid-HZ of \object{Kapteyn's Star} and close to the mean distance of Kapteyn b (see previous section). We have included Proxima Cen in the table even though it is a M5.5 V star ($R = 0.145 R_{\odot}$ from interferometry -- \citet{seg03}), and much smaller than the other stars. It is included since it has well determined properties -- especially radius (determined via interferometry by \citet{seg03}), age and rotation. 

\begin{figure*}
\epsscale{1}
\plotone{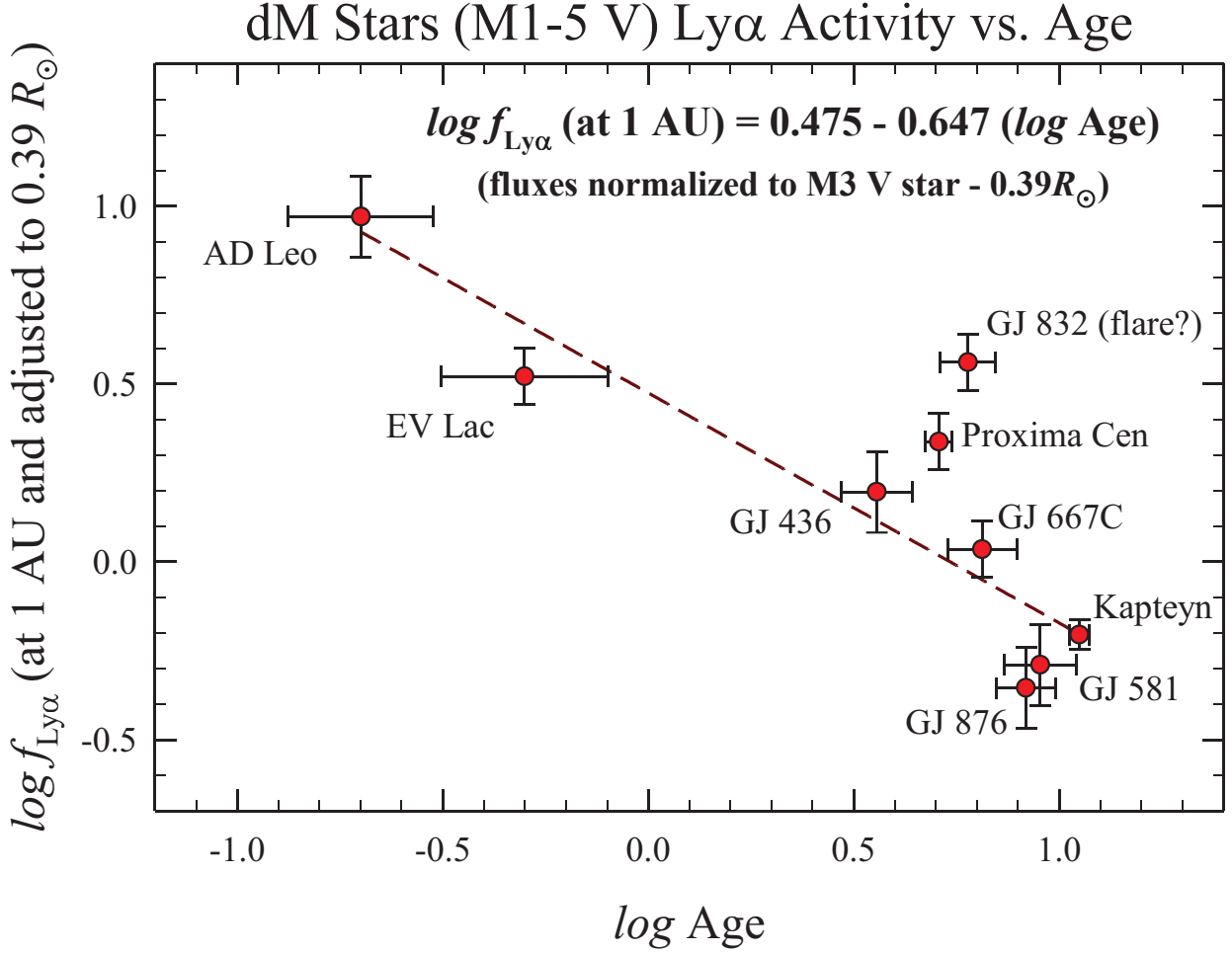}
\caption{The log of Ly$\alpha$ fluxes (at 1AU) for the M1--5 V program stars are plotted against the log of stellar age. The M1.5--M4 V stars in the plot, except Proxima Cen, have similar properties (except for age). Note that, as the Ly$\alpha$ emitting region is just above the stellar photosphere, it should theoretically scale with the size of the star. Therefore, we have scaled the emissions of all plotted stars to the same stellar radius of 0.39$R_{\odot}$ (M3 V star -- \citealt{rug15}). This plot shows the decline in Ly$\alpha$ flux with stellar age. A least squares fit to the data is shown and given by the relation: $log~f_{\rm Ly\alpha}~(1~{\rm AU}) = 0.475 - 0.647(log~Age)$. The decrease in Ly$\alpha$ flux over a nominal age range of 0.5 to 10.0 Gyr is $\sim$7/1. For comparison the corresponding decline in coronal X-ray emission (from Fig.~\ref{fig4}) over the same nominal age range of is $\sim$71/1.}
\label{fig8}
\end{figure*}

The integrated Ly$\alpha$ emission fluxes (at a reference stellar distance of 1 AU) for this small sample of red dwarfs are plotted against age in Fig.~\ref{fig8}, illustrating the decline in FUV activity with age. To better account for the range of spectral types (and hence Ly$\alpha$ emitting areas) in our sample, each star's flux has been adjusted to the radius of a M3 V star -- 0.39$R_{\odot}$ \citet{rug15}. For the dM stars in the plot, radii were determined based on spectral type \citep{rug15}, except for Proxima Cen and \object{Kapteyn's Star}, whose interferometric radii were used instead. Ly$\alpha$ is used as a proxy for FUV emissions as a whole. This is because, as discussed earlier, the Ly$\alpha$ flux makes up over $80-90$\% of the total FUV flux for dM stars. This estimate of Ly$\alpha$ flux contribution to the total EUV--FUV region is scaled from solar measures \citep[see][]{fra13,lin14}. The Sun is the only cool star with complete X-ray--EUV--FUV--NUV spectral coverage, due to nearly total H \textsc{i} absorption at $\sim$500 -- 900\AA.

As shown in Fig.~\ref{fig8}, the decay in the Ly$\alpha$ flux with age is less rapid than found for the coronal X-ray fluxes. Using the relation of \textit{log}~$f_{\rm Ly\alpha}$ (at 1 AU) = 0.475 -- 0.647(\textit{log} Age [Gyr]) given in the figure, the decline in the Ly$\alpha$ flux from young stars (reference age = 0.5 Gyr) to old disk/Pop II stars (reference age $\sim$10 Gyr) is $\sim$7/1. But it should be noted that the relation between Ly$\alpha$ flux and age is defined by only nine stars. In Fig.~\ref{fig8}, the high Ly$\alpha$ flux for the assumed age of GJ 832 could arise from uncertainties in the reconstructed Ly$\alpha$ profile, or from its age estimate. This star is plotted, but not used in the activity-age relations. Additional Ly$\alpha$ emission measures with \textit{HST-COS} are sorely needed for stars with reliable ages. A similar decline in chromospheric Ca \textsc{ii} \textit{HK} emission for red dwarfs has been found by \citet{egg90}. Using our previous reference ages of $\sim$0.5 Gyr and 10 Gyr for young and old stars, respectively, a decrease in Ca \textsc{ii} \textit{HK} emission flux is $\sim$15/1. \citet{mam08} find a decrease in Ca \textsc{ii} \textit{HK} flux over the same age range for K stars is about 25/1. Whereas a smaller decline of H$\alpha$ emission has been recently reported by \citet{wes15} for M dwarfs.




\section{Discussion \& Summary}

\object{Kapteyn's Star} is an exceptional star. It is one of the oldest stars in our red dwarf sample and, at a distance of 12.76 light years, the closest Pop II star to the Sun. As such, it serves as an important test bed for stellar dynamos and for the resulting coronal and chromospheric XUV emissions of slowly rotating, metal-poor, near-fully convective red dwarfs. The star also serves as an old-age anchor for red dwarf Activity-Rotation-Age relations. Importantly, \object{Kapteyn's Star} is a surrogate for studying the XUV irradiances of older red dwarfs and the effects of these emissions on hosted HZ planets.

\noindent$\bullet$ Activity of \object{Kapteyn's Star}:

It is interesting that a star as old as \object{Kapteyn's Star} still has manifestations of magnetic activity that are comparable to (or even stronger than) our Sun. These include star spots (from our photometry), chromospheric emissions (e.g. Ly$\alpha$, Mg \textsc{ii} \textit{hk} and Ca \textsc{ii} $HK$, as shown in Figs.~\ref{fig5} \& \ref{fig6}) and Transition Region FUV emissions -- such as N \textsc{v}, C \textsc{iv}, Si \textsc{iv}, He \textsc{ii} (see Fig.~\ref{fig5}).  High energy plasmas are also present, including variable coronal X-ray emissions. Judging from the available Ca \textsc{ii} $HK$ measures, \object{Kapteyn's Star} also may be undergoing flares that could have impacts on the atmospheres of hosted planets. More observations (\textit{UBV} time-series photometry) and additional Ca \textsc{ii} spectroscopy are needed to verify this. As indicated in this study, the levels of resulting XUV emissions in the star's HZ are similar to (or somewhat higher) than the Earth receives from the Sun today. Unfortunately, there are no measures of mass loss (stellar wind) for a star as old as \object{Kapteyn's Star}. However, \citet{woo01} have measured the mass loss of the solar-age M5.5 V star Proxima Centauri to be \textit{\.{M}} $<$ $0.2$ \textit{\.{M}}$_\odot$. Scaling from this, the winds of \object{Kapteyn's Star} are expected to be of similar magnitude or less. Thus, at its present age and activity level, an Earth-like planet (such as the exoplanet candidate Kapteyn b) could probably support life, assuming its atmosphere (and any primitive life) survived the expected high level of activity when the star was young. However, as discussed previously and shown in Figs.~\ref{fig4} and~\ref{fig8}, the levels of X-ray and FUV emissions of \object{Kapteyn's Star} were most likely $\sim86\times$ and $\sim8\times$ higher, respectively, when the star was young ($\sim0.5$ Gyr), compared to the current activity at its present, adopted age of $\sim$11.5 Gyr. 

\noindent$\bullet$ XUV Irradiances of Red Dwarfs:

Relations are provided (Figs.~\ref{fig4} and \ref{fig8}) for X-ray and FUV Ly$\alpha$ irradiances (for convenience, fluxes referenced at 1 AU) for red dwarfs (M0--5 V) with ages from $\sim$100 Myr to 11.5 Gyr. Since the Ly$\alpha$ emission feature dominates the FUV region, it is a good proxy for FUV emissions as a whole. These relations are important for computing the effects of ionizing XUV radiation on the atmospheres of red dwarf HZ planets. These XUV irradiance values complement similar measures for solar-type stars given previously by \citet{rib05}. In particular, the strong XUV emissions of young stars (age $<$ 1 Gyr), when coupled with stellar winds, flares and possible CME events play crucial roles on the ability of a hosted HZ planet to retain life-supporting atmospheres and water inventories. If the planet's atmosphere survives this strong XUV irradiation (or is replenished later by outgassing and cometary collision), then that planet (at a later time) may have the potential to support the development of life. Of course, as discussed by \citet{gud15} and others, the properties of the planet itself -- such as its orbit, instellation, mass, atmosphere, composition, water inventory, albedo, rotation rate, plate tectonics and geomagnetic field also play important roles on the planet's suitability for life.

We note that the relations provided here cover M0--5 V stars. Cooler red dwarfs (later than M6 V) have not yet been extensively studied (in terms of activity-rotation-age) to the point where calibrations  can be furnished. The few M8--9 V stars with activity-rotation-age data show hints that these very low mass stars do not follow the relations that we have found for early to mid dM stars. In our limited data, it appears that dM7+ stars retain fast rotations, even at old ages.

\noindent$\bullet$ The H \textsc{i} Ly$\alpha$ Emission Feature:

As discussed in detail earlier, we have taken advantage of the large radial velocity ($RV = +245.2$ km s$^{-1}$) of \object{Kapteyn's Star} to secure a precise measure of the H \textsc{i} Ly$\alpha$ emission flux. As shown in Fig.~\ref{fig7}, this is possible since the long-wavelength portion of the stellar emission line is essentially free from H \textsc{i} ISM absorption. The Ly$\alpha$ emission is by far the strongest emission feature in the EUV--FUV spectral region of older red dwarfs. In the case of \object{Kapteyn's Star}, it is estimated that its flux contributes about 80--90\% of the total EUV--FUV (300 -- 1900\AA) flux. Thus, this strong feature serves as an excellent tracer of the full EUV--FUV irradiances for red dwarfs. For example, as given in Table 3, the Ly$\alpha$ flux is 550$\times$ stronger than the C \textsc{iv} 1550 emission and over a thousand times stronger than the emission line fluxes of C \textsc{ii} 1335 and Si \textsc{iv} 1400. But it should be noted that \object{Kapteyn's Star} has a low metal abundance.

\noindent$\bullet$ Red Dwarfs as Suitable Hosts for Habitable Planets:

One of the primary differences in the magnetic-related behavior of red dwarfs and solar-like stars is that red dwarfs have an extended, active youth and remain active over much longer times \citep{wes04,boc05}, while the X-ray coronal and UV chromospheric activity of solar-type stars, including large flares \citep[see][and references therein]{mae12,not13}, decrease more rapidly with age. The extended activity of red dwarfs (and accompanying strong XUV emissions and flares) could pose major impediments for the development of life on HZ planets such as Kapteyn b. For example, the atmospheres of close-in HZ planets (even with significant geo-magnetic fields) could be threatened by frequent large flares, and the probable accompanying coronal mass ejections, that could persist for a few billion years. This sustained activity of red dwarfs could erode and even remove the atmospheres of young HZ planets, making it less likely for life to be presented with the necessary environment to develop and evolve in. Another possible problem for the habitability of close-in HZ planets hosted by red dwarfs is tidal locking. Tidal-locking reduces the rotation period of the planet and possibly reduces its protective geomagnetic field. Without a robust magnetic field, ion-pick up mechanisms, combined with CME events, strong winds, and flares, could strip the planet of its atmosphere \citep{lam08}. However, \citet{dri15} recently showed that a tidally-locked planet could have its magnetic field sustained (or even enhanced) via the tidal heating of its geomagnetic-generating iron-nickel core.

\noindent$\bullet$ The Rotation of \object{Kapteyn's Star}, and Consequences for Kapteyn b:

The analysis of CCD photometry indicates the presence of small amplitude, periodic, quasi-sinusoidal variations assumed to arise from rotation with a period of $P_{\rm rot} = 82.2\pm11.3$ days. This period agrees well with the period of $P_{\rm rot} = 83.7\pm0.3$ days returned from the independent analysis of archival Ca \textsc{ii} \textit{HK} emission measures. We note that many of the older stars in our \textit{Living with a Red Dwarf} program have been found to have similar light variations as well as long rotation periods \citep[e.g. see][]{eng11}. This rotation period (if confirmed by additional observations) does not appear to be related to the orbital periods of the planets proposed by \citet{ang14,ang15}. It also does not agree with the rotation period of $143\pm3$ days proposed by \citet{rob15}, which lead them to question the existence of Kapteyn b. Thus, in an indirect way, our result lends support to the \citet{ang14,ang15} findings that \object{Kapteyn's Star} hosts at least two possible super-earth size planets. However, additional photometry (and spectroscopy) are still preferred to confirm the star's rotation period. Additional high precision radial velocity measures of the star would also be very beneficial.

\section{Future Plans}

We are continuing time-series photometry to verify the period that we have found. The photometry is being carried out by members of the \textit{Variable Stars South} (VSS) organization. From past experience with photometry of similar old, inactive red dwarfs, it may take a few more years to reliably recover long-period brightness variations.

We plan to request additional \textit{HST} spectra of \object{Kapteyn's Star} to search for possible changes in the Ly$\alpha$ emission flux that could arise from variable stellar winds, flares and a possible activity cycle. It will be informative to investigate whether such an old, inactive star has an activity cycle. Also, we have identified additional high-velocity K5--M5 V stars with large radial velocities ($RV > 200$ km s$^{-1}$). As in the case of \object{Kapteyn's Star}, it will be possible to secure precise measures of the Ly$\alpha$ emission fluxes essentially free from H \textsc{i} ISM absorption and geo-coronal emission for these stars. In cases where the star's radial velocity is blue-shifted, we will request \textit{HST-STIS} high resolution spectroscopy to search for evidence of bow shocks as the rapidly approaching star's winds and astrosphere clash with the ISM. We also plan to carry out photometry of these stars to determine their (expected long) rotation periods and fill-factors (percentage of surface coverage) for star spots. Photometry of some of these stars is already underway, to determine rotation periods. Also, subsequent \textit{Living with a Red Dwarf} program papers are being prepared for publication, detailing the Rotation, Age and XUV activity measures that cover a broad age range, from $\sim$50 Myr to $\sim$12 Gyr \citep[][Engle et al. 2016 in preparation]{eng15}.

As an end note, because of the large numbers of red dwarfs and their longevities (often exceeding 100 Gyr), planetary systems hosted by red dwarfs (like \object{Kapteyn's Star}, and billions of others in our Galaxy) may have a greater possibility and time to develop more advanced life than presently on Earth. The study of these stars, along with the determination of age calibration relations, provides a dynamic evolutionary history and an assessment of planetary system stability with specific attention towards exoplanetary atmospheres, habitability and life.



\acknowledgments

This research is supported by the \textit{NSF} and \textit{NASA} through grants \textit{NSF/RUI-1009903}, \textit{HST-GO-13020.001-A} and \textit{Chandra} Award \textit{GO2-13020X} to Villanova University, and we are very grateful for this support. We also wish to acknowledge the availability of \textit{HST} data through the \textit{MAST} website hosted by the \textit{Space Telescope Science Institute} and the \textit{Chandra} X-ray data through the \textit{HEASARC} website hosted by the Astrophysics Science Division at \textit{NASA/GSFC} and the High Energy Astrophysics Division of the Smithsonian Astrophysics Observatory (\textit{SAO}). 

We would also like to thank an anonymous referee for the useful suggestions for improving the manuscript. We are also grateful for observing time on \textit{PROMPT} from the University of North Carolina at Chapel Hill, provided via funding from the \textit{Robert Martin Ayers Sciences Fund}. This research made use of public databases hosted by SIMBAD, and maintained by CDS, Strasbourg, France.




{\it Facilities:} \facility{XMM-Newton}, \facility{Chandra}, \facility{HST (COS)}.



\begin{deluxetable}{ccc}
\tabletypesize{\small}
\tablecaption{Selected Properties of \protect\object{Kapteyn's Star} \label{tbl1}}
\tablehead{\colhead{Quantity} & \colhead{Measurement} & \colhead{Source} \\ 
\colhead{} & \colhead{} & \colhead{} } 
\startdata
\textit{V}-mag & $8.853\pm0.008$ & SIMBAD \& this study \\
\textit{B-V} & $1.57\pm0.012$ & SIMBAD \\
Sp Type & sdM1.0 (or M1.5 V) & \citet{giz97} \\
Distance (pc) & $3.91\pm0.01$ & \citet{van07} \\
$T_{\rm eff}$ (K) & $3570\pm80$ & \citet{ang14} \\
$log~g$ & $4.96\pm0.13$ & \citet{seg03} \\
$L/L_{\odot}$ & $0.012$ & \citet{ang14} \\
Radius ($R_{\odot}$) & $0.291\pm0.025$ & \citet{seg03} \\
Mass ($M_{\odot}$) & $0.281\pm0.014$ & \citet{seg03} \\
$[M/H]$ & $-0.86\pm0.05$ & Woolf \& Wallerstein 2005 \\
$v~sin~i$  (km/s) & $< 3.0$ km s$^{-1}$ & \citet{ang14} \\
$P_{\rm rot}$ (d)  & $82.2\pm11.3$  & This study (Phot) \\
                   & $84.7\pm0.5$    & This study (Ca \textsc{ii})\\
                   & 143 days & \citet{rob15} \\
Age (Gyr) & $11.5^{+0.5}_{-1.5}$ Gyr & \citet{kot05} / \citet{wyl10} \\
 &  &  \\
$L_{\rm X}$($0.3-2.5$ keV) & $2.4-6.0\times10^{26}$ ergs s$^{-1}$ & This study \\
$\langle f_{\rm X} \rangle~({\rm at}~1~{\rm AU})$         & $0.157$ erg s$^{-1}$ cm$^{-2}$     & This study \\
$\langle L_{\rm X} / L_{\rm bol}\rangle$  &  $9.53\times10^{-6}$  &  This study  \\
$f_{\rm Ly\alpha}~({\rm at}~1~{\rm AU})$  & $0.347$ erg s$^{-1}$ cm$^{-2}$     & This study \\
Ca \textsc{ii} \textit{HK} $\langle S \rangle$ & $\sim0.24-0.28$      & From \citet{ang14} \\
Habitable Zone (HZ) & $0.13-0.24$ AU & \citet{ang14} \\
S/S$_{\oplus}$\tablenotemark{a} & 0.40 (at 0.17AU(Kapteyn b)) \\
 & 0.12 (at 0.31AU(Kapteyn c)) & \citet{ang14} \\
\enddata
\tablenotetext{a}{ratio of total instellation at 0.17 AU relative to insolation (solar flux) at 1 AU}




\end{deluxetable}

\begin{deluxetable}{ccccc}
\tabletypesize{\small}
\tablecaption{X-ray Properties of \protect\object{Kapteyn's Star} \label{tbl2}}
\tablehead{\colhead{Parameter} & \colhead{ROSAT 1T} & \colhead{ROSAT 1T} & \colhead{Chandra 2T5} & \colhead{Chandra 2T10} \\ 
\colhead{Value} & \colhead{Mar/Apr 1992} & \colhead{Sep/Oct 1992} & \colhead{} & \colhead{} } 
\startdata
$N_{\rm H}$ &  &  & \multicolumn{2}{c}{$8\times10^{17}$} \\
$kT_{\rm 1} (keV)$ & $0.09-0.22$ & $0.06\pm0.31$ & 0.07 & 0.07 \\
relative normalization  &  &  & 0.00640 & 0.015 \\
$kT_{\rm 2} (keV)$ &  &  & 0.63 & 0.62 \\
relaitve normalization  &  &  & 0.00013 & 0.00014 \\
$f_{\rm X}$ (erg s$^{-1}$ cm$^{-2}$) & $1.33\times10^{-13}$ & $3.08\times10^{-13}$ & $2.30[1.1]\times10^{-13}$ & $3.30[1.0]\times10^{-13}$ \\
$L_{\rm X}$ (erg s$^{-1})$ & $2.43\times10^{26}$ & $5.63\times10^{26}$ & $4.21[2.01]\times10^{26}$ & $6.04[1.83]\times10^{26}$ \\
$log~L_{\rm X}$ (erg s$^{-1})$ & 26.39 & 26.75 & $26.62[0.17]$ & $26.78[0.12]$ \\
 &  &  & \multicolumn{2}{c}{$\langle L_{\rm X} \rangle = 5.125[2.718]\times10^{26}$} (erg s$^{-1}$)\\
 &  &  & \multicolumn{2}{c}{$\langle log~L_{\rm X} \rangle = 26.71[0.21]$} (erg s$^{-1}$)\\
\enddata
\end{deluxetable}

\begin{deluxetable}{cccccc}
\rotate
\tabletypesize{\small}
\tablecaption{FUV--UV Emissions of \protect\object{Kapteyn's Star} \label{tbl3}}
\tablehead{\colhead{Emission Feature} & \colhead{flux} & \colhead{flux} & \colhead{Ratio (Ly$\alpha$} & \colhead{flux} & \colhead{flux (Kap. HZ) /} \\ 
\colhead{} & \colhead{Received} & \colhead{(at 1 AU)} & \colhead{/ feature)} & \colhead{(at 0.17 AU)} & \colhead{flux (Earth pos.)} \\
\colhead{} & \colhead{(erg s$^{-1}$ cm$^{-2}$ \AA$^{-1}$)} & \colhead{(erg s$^{-1}$ cm$^{-2}$ \AA$^{-1}$)} & \colhead{} & \colhead{(erg s$^{-1}$ cm$^{-2}$)} & \colhead{}} 
\startdata
H \textsc{i} Ly$\alpha$ 1215.67\AA & $5.32\times10^{-13}$ & 0.347 & 1.0 & 11.99 & $1.71\times$ \\
C \textsc{ii} 1335\AA & $2.82\times10^{-16}$ & $1.83\times10^{-4}$ & 1927.8 & 0.0063 &  \\
Si \textsc{iv} 1400\AA & $5.27\times10^{-16}$ & $3.43\times10^{-4}$ & 1011.7 & 0.012 &  \\
C \textsc{iv} 1550\AA & $9.69\times10^{-16}$ & $6.30\times10^{-4}$ & 550.8 & 0.022 &  \\
Mg \textsc{ii} \textit{hk} 2800\AA & $2.78\times10^{-14}$ & 0.018 & 19.28 & 0.626 &  \\
X-ray ($0.3-2.5$ keV) \\
($5.0-62$\AA) & $2.40\times10^{-13}$ & 0.156 & 1.86 & 5.408 & $4.10\times$ \\
\multicolumn{6}{c}{Sun (mean)} \\
H \textsc{i} Ly$\alpha$ 1215.67\AA & 7.00\tablenotemark{a} & 1.0 &  &  \\
X-ray ($0.2-2.5$ keV) & 1.32\tablenotemark{b} & 5.3 &  &  \\
\enddata
\tablenotetext{a}{from Linsky et al. (2014)}
\tablenotetext{b}{Computed from Ayres (2014)}
\end{deluxetable}

\begin{deluxetable}{cccccccc}
\tablecaption{The Integrated Fluxes of Red Dwarfs Possessing Ly$\alpha$ Measures 
\label{tbl4}
}

\tablehead
{
\colhead{Star} & 
\colhead{Sp.} & 
\colhead{$f_{\rm{Ly}\alpha}$} & 
\colhead{$f_{\rm{Ly}\alpha}$} & 
\colhead{$f_{\rm{Mg} \textsc{ii}}$} & 
\colhead{Age} & 
\colhead{Rotation} & 
\colhead{Age} \\
 
\colhead{} & 
\colhead{Type} & 
\colhead{(1 AU)} & 
\colhead{(0.17 AU)} & 
\colhead{(1 AU)} & 
\colhead{(Gyr)} & 
\colhead{(days)} & 
\colhead{Ref.}
} 

\startdata
AD Leo  & M3.5 V          & 9.33  & 322.8  & 2.19                   & $\sim0.2\pm0.1$      & $2.23\pm0.006$                  & f \\
EV Lac  & M3.5 V          & 3.07  & 106.2  & 0.637\tablenotemark{a} & $\sim0.4\pm0.1$          & $4.35\pm0.001$\tablenotemark{b} & g \\
GJ 436  & M3.0 V          & 1.571 & 54.36  & 0.104                  & $3.6\pm0.8$          & --                              & h \\
Proxima & M5.5 V          & 0.301 & 10.42  & 0.045\tablenotemark{a} & $5.1\pm0.5$          & $82.9\pm0.6$                    & i \\
GJ 832  & M1.5 V          & 5.17  & 178.89 & 0.311                  & $6\pm1.5$            & --                              & h \\
GJ 667C & M1.5 V          & 1.54  & 53.29  & 0.113                  & $6.5\pm1.4$          & $103\pm2$\tablenotemark{c}      & j \\
GJ 876  & M4  V           & 0.409 & 14.15  & 0.032                  & $7.9\pm1.2$          & $116.4\pm2.5$\tablenotemark{d}  & j \\
GJ 581  & M2.5 V          & 0.513 & 17.75  & 0.036                  & $9\pm2$              & $157.6\pm2.5$\tablenotemark{d}  & k \\
Kapteyn & M1.5 V          & 0.347 & 11.99  & 0.018                  & $11.5^{+0.5}_{-1.5}$ & $83.7\pm0.3$\tablenotemark{e}   & l \\
\enddata

\tablenotetext{a}{\citet{doy90}} 
\tablenotetext{b}{\citet{dal11}}
\tablenotetext{c}{Analysis of HARPS Ca \textsc{ii} \textit{HK} data from \citet{ang14}}
\tablenotetext{d}{Ground-based photometry carried out by us}
\tablenotetext{e}{This Paper}
\tablenotetext{f}{Castor Moving Group, from \citet{klu14}}
\tablenotetext{g}{Possible UMa Moving Group, from \citet{klu14}}
\tablenotetext{h}{Age determined from our X-ray Activity-Age relationship}
\tablenotetext{i}{$\alpha$ Cen ABC triple system, see \citet{rei12} and references therein}
\tablenotetext{j}{Age determined from our Rotation-Age and X-ray Activity-Age relationships}
\tablenotetext{k}{Age determined from our Rotation-Age and X-ray Activity-Age relationships, considering that the star's metallicity and space motions are inconsistent with the Halo population, putting an upper limit on the final age}
\tablenotetext{l}{Age determined from our Rotation-Age and X-ray Activity-Age relationships, along with metallicity and space motions indicative of a Halo population star}

\end{deluxetable}

\clearpage
\nocite{*}
\bibliography{kapteyn_ApJ}

\clearpage

\end{document}